\documentclass[a4paper,11pt]{article}
\pdfoutput=1 
\usepackage{jcappub}
\usepackage{xcolor}
\usepackage{lineno}
\usepackage{soul}
\usepackage[normalem]{ulem}
\usepackage[T1]{fontenc} 
\usepackage{orcidlink}

\newcommand{\CH}[1]{{\color{black}{{#1}}}}

\title{\boldmath Bispectrum BAO and the baryon-dark matter relative velocity}

\author[a,b]{Adriana Nadal-Matosas\orcidlink{0009-0007-5871-2587},}
\author[a,b,c]{Héctor Gil-Marín\orcidlink{0000-0003-0265-6217},}
\author[a,d]{Licia Verde\orcidlink{/0000-0003-2601-8770}}

\affiliation[a]{Institut de Ciències del Cosmos (ICCUB), Universitat de Barcelona (UB), c.  Martí i Franquès, 1, 08028 Barcelona, Spain.}
\affiliation[b]{Departament de F\'{\i}sica Qu\`{a}ntica i Astrof\'{\i}sica, Universitat de Barcelona, Mart\'{\i} i Franqu\`{e}s 1, E08028 Barcelona, Spain}
\affiliation[c]{Institut d'Estudis Espacials de Catalunya (IEEC), Edifici RDIT, Campus UPC, 08860 Castelldefels (Barcelona), Spain}
\affiliation[d]{Institució Catalana de Recerca i Estudis Avançats (ICREA), Pg. Lluís Companys 23, E-08010 Barcelona, Spain}

\emailAdd{anadal@icc.ub.edu}
\emailAdd{hectorgil@icc.ub.edu}
\emailAdd{liciaverde@icc.ub.edu}

\abstract{We evaluate the Baryon Acoustic Oscillation (BAO) signal in the bispectrum as a tool to detect and characterize the relative velocity effect. We extend the existing framework by presenting an updated model for the redshift-space tree-level bispectrum that comprehensively incorporates all relative velocity terms. We introduce a novel, unbiased technique to extract the isotropic BAO dilation parameter ($\alpha_{\rm iso}$) solely from the bispectrum monopole. Validated against N-body simulations, this template-based extraction successfully recovers the acoustic scale and enhances the statistical constraining power by \CH{ $\sim22\%$ for a volume of 500($h^{-1}$Gpc)$^3$} , when analyzed in tandem with the pre-reconstruction power spectrum, offering a powerful complement to standard post-reconstruction pipelines. We quantify how individual relative velocity components distort both two- and three-point statistics. We find that these effects induce distinct systematic shifts in the extracted $\alpha_{\rm iso}$ between the two probes. We find systematic discrepancies of up to $2\%$ for $b_{v^2}=\pm0.05$ and up to  $20\%$ for $b_{\delta^{bc}}\le-2$. This differences demonstrate that a direct comparison of independent power spectrum and bispectrum BAO measurements can break parameter degeneracies and isolate the amplitude of these biases. Finally, we provide a concrete prescription to detect and constrain the three associated relative velocity bias parameters, showing that a joint analysis is highly sensitive to the $b_{v^2}$ and $b_{\delta^{bc}}$ amplitudes.
This establishes the bispectrum BAO as a robust cosmological probe for current and next-generation galaxy surveys, serving both as a cross-check for standard analyses and a crucial diagnostic tool against systematic biases.}

\begin{document}
\maketitle
\flushbottom

\section{Introduction}
\label{Sec:Introduction}

The Baryon Acoustic Oscillation (BAO) signal imprinted in the large-scale distribution of galaxies represents one of the main probes of cosmology today. Increasingly large and high-quality datasets provide unprecedented statistical power to constrain cosmological parameters and, in general, test the standard cosmological model \cite{Eisenstein_2007_BAO,Blake2003,Seo2003}. BAO measurements from two-point statistics, such as the power spectrum or the correlation function, have played a key role in measuring cosmological parameters from massive spectroscopic galaxy surveys over the past decade \cite{Boss,eBossBAO,DESIDR1,DESIDR2}. 
With larger survey volumes driving down statistical errors, meticulous control of systematic errors is a must. Effects that were previously neglected as subdominant errors now become relevant, potentially becoming independent cosmological probes.

As modern galaxy surveys explore volumes substantially larger than those of previous generations, they provide sufficient statistical power to exploit information beyond traditional two-point statistics. The bispectrum is the leading higher-order statistic of the galaxy density field, probing three-point correlations in Fourier space. It is sensitive to non-Gaussian features of the matter distribution arising from non-linear gravitational evolution, non-linear galaxy bias, and potentially primordial non-Gaussianity. Consequently, the bispectrum provides information complementary to that contained in the power spectrum and can help to break degeneracies among cosmological parameters.
In recent years, the bispectrum has become an increasingly important component of the state-of-the-art large-scale structure analyses \cite{Scoccimarro2001,HGM2015a,HGM2015,HGM2017,Philcox2022,Ivanov2023,Damico2024,SNM2025,SNM2026}. However, fully exploiting its cosmological information content remains a challenge. The application of the bispectrum to observational data is computationally demanding due to the large size of the data vector, the strong correlations between its elements, the non-diagonal nature of its covariance matrix, and the complexity of accurately modeling both the signal and observational systematics. As a result, despite substantial theoretical and methodological developments, the use of the bispectrum for cosmological inference from real survey data remains comparatively less mature than that of two-point statistics.

Extracting the BAO signal from the bispectrum serves as an intermediate approach. While it does not fully exploit the complete information content of the signal, it enables a robust use of this higher-order statistic. 
By focusing on the BAO feature, the analysis is less susceptible to full-shape modeling systematics,
as flexible nuisance parameters can absorb, at least in part, incorrect modeling of the non-linear broadband.
As long as the underlying BAO structure is correctly captured, the bispectrum measurement can help tighten constraints and potentially break degeneracies affecting the standard 2-point function or power spectrum BAO measurements and associated systematics.

A promising application is the investigation of the relative velocity effect in galaxy clustering. The success of BAO analyses relies on accurately modeling how primordial acoustic oscillations, imprinted in the baryon–dark matter fluid at high redshift, propagate into the observed galaxy distribution through structure formation. This same process that sources the BAO results in a baryon–dark matter relative velocity that becomes supersonic after recombination \cite{Hirata2010}. 
This relative velocity has been shown to modulate early galaxy formation on scales comparable to the sound horizon \cite{Hirata2011}. The late-time large-scale structure may retain some memory from this relative velocity, potentially introducing a bias in BAO-based distance measurements \cite{Yoo2011,Dalal2010,Slepian2015,Blazek2016}. 
In this context, the bispectrum has been shown to be an effective tool in removing the relative velocity contribution in the power spectrum \cite{Yoo2011}.

The effect induced by velocity bias is well-motivated theoretically and confirmed by simulations \cite{Tseliakhovich2011,Visbal2012,Popa2016,Conaboy2023}, so it is expected to be present at some level, but is not the only process that could potentially be studied with a joint BAO power spectrum-bispectrum analysis.
Other proposed models introduce analogous oscillatory features on the clustering of dark matter and galaxies that can interfere with the BAO in a similar way. 
These include dark sector models, resulting in dark matter oscillations \cite{DAO,DAO2,DAO3,DAO4,DAO5}, and dark matter-baryon scattering effects \cite{scattering4,scattering2,scattering3,scattering5,scattering}. Also inflationary models introducing sharp and logarithmic features \cite{Features,Features2,SharpFeatures,logFeatures} that modulate similar scales to the BAO. Initial conditions modifications such as isocurvature and compensated isocurvature perturbations \cite{Isocurv,Isocurv2,CompIsocurv,CompIsocurvPlanck} have also been shown to result in a similar effect. While some of these hypotheses are purely speculative, investigating the effects of relative velocity may provide insights into them as well.

It is important to note that the standard (and state-of-the-art) BAO data analysis does not model the effect of relative velocities (or any other physical process with a similar effect on the BAO); it assumes that the effect is absent. This means that if the effect is present, even at a low level, it may bias the cosmological interpretation of the BAO signal. 

In this work we aim to investigate the potential of the bispectrum BAO measurement to detect the relative velocity effect. 
\CH{The objective is not a fully Bayesian constrain on the relative velocity bias parameters. Rather it is to explore how the measurement of the BAO in the bispectrum can help uncover and mitigate biases induced in the power spectrum only BAO analysis.}

The structure of the paper is as follows. In section \ref{Sec:RELVEL} we describe the modeling of the relative velocity in galaxy clustering. In section \ref{sec:BAO} we present the BAO analysis, introducing a novel methodology to extract the BAO signal from the bispectrum monopole. We present our results in section \ref{sec:Results} and conclude in \ref{Sec:Conclusions}.
\section{Relative Velocity Effect in Galaxy Clustering}
\label{Sec:RELVEL}

After cosmological recombination, baryons decoupled from the photon fluid and their sound speed dropped fast from $c_s \simeq 1/\sqrt{3}$ to $c_s \sim 10^{-5}$, generating a coherent supersonic relative velocity between the baryonic and cold dark matter (CDM) fluids \cite{Hirata2010}. This baryon-CDM streaming velocity, $v_\mathrm{bc}$, has a root-mean-square amplitude of approximately $30\,\mathrm{km\,s^{-1}}$ at $z \sim 1000$ and decays as $v_\mathrm{bc} \propto a^{-1}$ where $a$ is the scale factor. Despite this decay, the effect remains dynamically important at high redshifts, suppressing gas accretion into low-mass dark matter halos of characteristic mass $M_h \sim 10^6\,M_\odot$ and delaying the onset of early star and galaxy formation. Physically, the streaming velocity introduces a relative density perturbation \CH{$\delta_\mathrm{bc}=\delta_b(\mathbf{x})-\delta_c(\mathbf{x})$ and velocity divergence between the two fluids $\theta_\mathrm{bc}=\theta_c(\mathbf{x})-\theta_c(\mathbf{x})$}, which suppresses the formation of early-time halos at the Jeans scale ($k_J \sim 200\,\mathrm{Mpc}^{-1}$) \cite{Hirata2010,Hirata2011}.

These early differential conditions on low-mass halos can be transmitted to the low-redshift galaxy population through feedback processes associated with reionization, metal enrichment or supernova rates \cite{35,36,38,39,40,41,42}. Although these effects have been extensively studied, the exact significance on the late-time distribution of galaxies is difficult to predict from theory alone. Therefore, it is customary to assume that the amplitude of the additional terms sourced by the relative baryon-CDM sector is modulated using three bias parameters to be constrained by data \cite{Yoo2013,Barkana2011,Slepian2018,Beutler2017}. 

The large-scale galaxy number density is modified by the local amplitude of the streaming velocity, $v_\mathrm{bc}^2$, through the bias parameter $b_{v^2}$, as well as by the relative density contrast $\delta_\mathrm{bc}$ and the relative velocity divergence $\theta_\mathrm{bc}$, through the bias parameters $b_\delta^{\rm bc}$ and $b_\theta^{\rm bc}$, respectively. While $b_{v^2}$ modulates the overall halo abundance on large scales, $b_\delta^{\rm bc}$ and $b_\theta^{\rm bc}$ encode the response of the galaxy field to the relative-fluid perturbations seeded at recombination. Since all three effects are coherent on the BAO scale ($\sim 150\,\mathrm{Mpc}$), being sourced by the same pre-recombination acoustic dynamics, a nonzero value of any of these bias parameters, if unaccounted for, can systematically shift the inferred BAO peak position \cite{Dalal2010,Yoo2011}. 
With current surveys achieving sub-percent precision on $\alpha_{\rm iso}$ \cite{DESIDR2}, up-to-date analyses show this effect constitutes a potentially significant source of systematic uncertainty. Measurements show the resulting bias can reach $\sim1\%$ at the boundary of the $95\%$ confidence interval \cite{Beutler2017}, indeed comparable to the statistical budget of next-generation surveys. 

To study the potential bias, we start by modeling the relative effect on the distribution of galaxies. The galaxy density field is expanded to quadratic order in perturbations as follows \cite{Blazek2016,Schmidt2016}
\begin{align}
\label{eq:density}
\delta_g^{\,s}(\mathbf{x})&= b_1\,\delta_m(\mathbf{x})+ \frac{1}{2} b_2\left[\delta_m^2(\mathbf{x}) - \langle \delta_m^2 \rangle\right]+ \frac{1}{2} b_{s^2}\left[s^2(\mathbf{x}) - \langle s^2 \rangle\right]
\nonumber\\&\quad
+ b_{v^2}\left[v_{\rm bc}^2(\mathbf{x}) - \langle v_{\rm bc}^2 \rangle\right]+ b_\delta^{\rm bc}\delta_{bc}(\mathbf{x}) + b_\theta^{\rm bc}\theta_{bc}(\mathbf{x}) + \cdots ,
\end{align}

where $\delta_m(\mathbf{x})$ is the matter density field and $s(\mathbf{x})$ is the tidal tensor field given by $s_{ij}(\mathbf{x}) = \left( \nabla_i \nabla_j \nabla^{-2} - \frac{1}{3} \delta_{ij} \right) \delta(\mathbf{x})$. The bias parameters are $b_1$ the linear bias, $b_2$ the non-linear bias and $b_{s^2}$ the non-local bias. The relative density field $\delta_{\mathrm{bc}}$ describes the variation in the ratio of cold dark matter to baryons arising from different initial conditions after decoupling. The relative velocity and divergence fields, $v_{\rm bc}$ and $\theta_{\mathrm{bc}}$, characterize the same effect in the velocity field.
Their transfer functions map this effect to the late-time large-scale structure (LSS) and are defined as
\begin{align}
    T_{\mathrm{bc}}(k)=\frac{T_b(k)-T_{\mathrm{cdm}}(k)}{T_m(k)}, \quad T_v(k)\propto\frac{T_{v,b}(k)-T_{v,\mathrm{cdm}}(k)}{T_m(k)},
\end{align}
where $T_b$ and $T_{\mathrm{cdm}}$ are the density transfer functions of baryons and cold dark matter and $T_{v,b}$ and $T_{v,\mathrm{cdm}}$ their respective velocity transfer functions. 

The velocity transfer functions are related to the density ones following $T_{v,i}=-a\dot{T}_i/k$. We obtain all transfer functions from the Boltzmann solver  \textsc{Class}\footnote{\href{http://class-code.net}{http://class-code.net}} \cite{Class11}. The normalization of the velocity transfer function at any redshift is obtained following
\begin{align}
\sigma_{v_{\mathrm{bc}}}^2&=\int_0^{\infty} \frac{k^2\,dk}{2\pi^2}\,T_v^2(k)\,P_{\mathrm{lin}}(k),
\end{align}
with $P_{\mathrm{lin}}(k)$ the linear matter power spectrum. Note we adopt the nomenclature from \cite{Slepian2015}, where our normalization $\sigma_{v_{\mathrm{bc}}}$ is a factor $\sqrt{3}$ times smaller than the one defined in \cite{Yoo2011}.

\begin{figure}[ht]
    \centering
    \includegraphics[width=\textwidth]{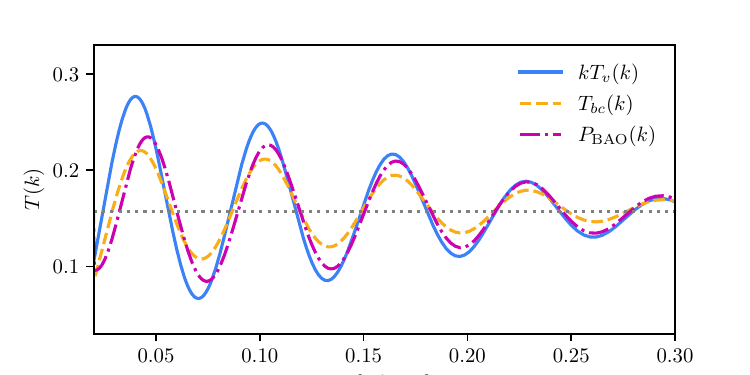}
    \caption{Oscillatory features and phase shifts in transfer functions. Shown is a comparison of the relative density transfer function $T_{bc}(k)$ (dashed yellow), the scaled relative velocity transfer function $kT_{v}(k)$ (solid blue), and the "wiggle" component of the power spectrum $P_{\rm BAO}(k)$ (dash-dotted magenta). The three functions have been arbitrarily rescaled for easier comparison. The observable phase shift between these curves, if unmodeled, represents a potential source of systematic bias in BAO scale determination.}
    \label{fig:oscillations}
\end{figure}

Figure \ref{fig:oscillations} compares the oscillatory features found in the relative density transfer function, $T_{bc}(k)$, the scaled relative velocity transfer function, $kT_{v}(k)$, and the wiggle present in the power spectrum, $P_{\rm BAO}(k)=P(k)/P_{\rm nowiggle}(k)$. The figure shows that the different features are present in the same wavenumber range as the BAO. It also highlights a relative phase shift of the oscillations, which, if unaccounted for,  can be a potential source of systematic error for the BAO measurement. In this work, we model the clustering signal of galaxies including the relative velocity effect and explore the bias that results when it's not taken into account during the BAO analysis. 

\subsection{Power Spectrum Velocity Bias model}
For modeling the power spectrum, we adopt the framework proposed in \cite{Beutler2017}, which builds on the models of \cite{McDonaldRoy2009}, \cite{TNS2010}, and \cite{Saito2014}, and extends it to include the relative velocity terms of \cite{Yoo2011} and \cite{Schmidt2016} in redshift space. The density field from equation \ref{eq:density} results in the following power spectrum structure,
\begin{align}
\label{eq:pk_datavector}
P_g(k,\mu)&= P_{g,\mathrm{NL}}(k,\mu)+ b_{v^2}\Big[b_1 P_{\delta|v^2}(k)+ b_2 P_{\delta^2|v^2}(k)+ b_s P_{s^2|v^2}(k)+ b_{v^2} P_{v^2|v^2}(k)\Big]
\nonumber\\&\quad
+ b_1 b_{v^2} P_{\mathrm{adv}|\delta}(k)+ 2 b_1 b_\delta^{\mathrm{bc}} P_{\delta|\delta_{\mathrm{bc}}}(k)+ 2 b_1 b_\theta^{\mathrm{bc}} P_{\delta|\theta_{\mathrm{bc}}}(k) \nonumber\\&\quad
- 2 f \mu^2\Big[b_{v^2}\big(b_1 P_{\delta|v^2 v_\parallel}(k)+ P_{\mathrm{adv}|v_\parallel}(k)\big)- b_\theta^{\mathrm{bc}} P_{\delta|\theta_{\mathrm{bc}}}(k)+ b_\delta^{\mathrm{bc}} P_{\delta|\delta_{\mathrm{bc}}}(k) \nonumber\\&\qquad\qquad
+ b_{v^2}\big(P_{v^2|v_\parallel}(k)+ P_{v^2|\delta v_\parallel}(k)\big)\Big]\nonumber\\&\quad+ f^2 \mu^4\, b_{v^2}\,P_{v_\parallel|v^2 v_\parallel}(k)- f^2 \mu^2\, b_{v^2}\left[I_1(k) + \mu^2 I_2(k)\right],
\end{align}
where the terms involving \CH{two powers of $b_\delta^{\mathrm{bc}}$ and $b_\theta^{\mathrm{bc}}$} are expected to be an order of magnitude smaller and therefore omitted \cite{Schmidt2016}. 
The redshift space galaxy non-linear power spectrum, $P_{g,\mathrm{NL}}$, is computed following 
\begin{align}
P_{g, \rm NL}(k, \mu) &= D_{\mathrm{FoG}}^P\left(k, \mu, \sigma_{\mathrm{FoG}}^P[z]\right) \Big[ 
P_{g,\delta\delta}(k) + 2f\mu^2 P_{g,\delta\theta}(k) + f^2\mu^4 P_{\theta\theta}(k) \notag \\
&\quad + b_1^3 A(k, \mu, f/b_1) + b_1^4 B(k, \mu, f/b_1) \Big].
\end{align}
Here $P_{g,\delta\delta}$, $P_{g,\delta\theta}$ and $P_{\theta\theta}$ are the density and divergence auto- and cross-correlations. We compute their non-linear SPT 1-loop predictions using the code \textsc{PTcool}\footnote{\href{https://github.com/hectorgil/PTcool}{https://github.com/hectorgil/PTcool}}. The expressions corresponding to these and every relative velocity term can be found in Appendix \ref{sec:appendix_A}. 

The terms driving a possible bias in the BAO measurement and interpretation are those with oscillatory features that are out of phase with the power spectrum BAO signal.
These terms are $P_{\delta|v^2}(k)$, $P_{\mathrm{adv}|\delta}(k)$, $P_{\delta|\delta_{\mathrm{bc}}}(k)$ and $P_{\delta|\theta_{\mathrm{bc}}}(k)$, with their RSD counterparts. A comparison of these terms can be found in figure \ref{fig:integrals_pk} in Appendix \ref{sec:appendix_A}. For a more detailed study on these, we refer the reader to \cite{Schmidt2016,Beutler2017}. 

The redshift space power spectrum multipoles can be obtained by integrating over their dependence on the angle with respect to the line of sight,
\begin{equation}
    P_{g}^{(\ell)}(k) = \frac{2\ell + 1}{2} \int_{-1}^{+1} d\mu \, P_{g}(k, \mu) \mathcal{L}_{\ell}(\mu),
\end{equation}
where $\mathcal{L}_{\ell}(\mu)$ are the Legendre polynomials. In this work, we focus on the monopole ($\ell=0$) and quadrupole ($\ell=2$), the first non-zero multipoles where the majority of the signal is contained.

\subsection{Bispectrum Velocity Bias  Model}
For the bispectrum, we extend the work of \cite{Yoo2011}, which derived the tree-level real space bispectrum. We compute and present the redshift space relative velocity contributions for the bispectrum, also including the $\delta^{bc}$ and $\theta^{bc}$ terms which were not present in previous works.

We start from the galaxy density field in equation \ref{eq:density} and construct the corresponding relative velocity kernels.
The redshift space tree-level contributions of the relative velocity effect give the following redshift space bispectrum
\begin{align}
\label{eq:Bispectrum_full}
B_g&(\mathbf{k}_1,\mathbf{k}_2,\mathbf{k}_3)=D_{\text{FoG}}^{B}(k_{1}, k_{2}, k_{3}, \sigma_{\text{FoG}}^{B}) \left[2 Z_1(k_1)Z_1(k_2)Z_2(\boldsymbol{k}_1,\boldsymbol{k}_2)P_m(k_1) P_m(k_2) + {\rm cycl.} \right]
\end{align}

where the linear redshift space kernel, adding the relative density and divergence fields contributions, reads:
\begin{equation}
Z_1(k) =  b_1 + b_\delta^{bc} T_{bc}(k) + b_\theta^{bc} \frac{\sigma_{vbc}}{H_0} k T_v(k) + f\mu^2.
\end{equation}
The second-order redshift space kernel with the added relative density and divergence fields contributions is
\begin{align}
Z_2(\boldsymbol{k}_1, \boldsymbol{k}_2) = b_1 F_2&(\boldsymbol{k}_1, \boldsymbol{k}_2) + \frac{b_2}{2} + b_s S_2(\boldsymbol{k}_1, \boldsymbol{k}_2) + f\mu^2 G_2(\boldsymbol{k}_1, \boldsymbol{k}_2) \nonumber\\ &+ b_{v^2} \mu G_u(\boldsymbol{k}_1, \boldsymbol{k}_2) + \frac{f \mu k}{2} \left[ \frac{\mu_1}{k_1} Z_1(k_2) + \frac{\mu_2}{k_2} Z_1(k_1) \right].
\end{align}
With $\boldsymbol{k} = \boldsymbol{k}_1 + \boldsymbol{k}_2$, $k = |\boldsymbol{k}|$, and $\mu = \hat{\boldsymbol{k}} \cdot \hat{\boldsymbol{z}}$ (where $\hat{\boldsymbol{z}}$ is the line-of-sight direction).
In equation \ref{eq:Bispectrum_full}, a Fingers-of-God (FoG) term \CH{is added} to extend the range of this model into the mildly non-linear regime: 
\begin{equation}
    D_{\text{FoG}}^{B}(k_{1}, k_{2}, k_{3}, \sigma_{\text{FoG}}^{B}) = \left( 1 + [k_{1}^{2}\mu_{1}^{2} + k_{2}^{2}\mu_{2}^{2} + k_{3}^{2}\mu_{3}^{2}]^{2} (\sigma_{\text{FoG}}^{B})^{4}/2 \right)^{-2}\,.
\end{equation}

Similarly to the power spectrum, the redshift space anisotropic bispectrum can also be decomposed in spherical harmonics \cite{Scoccimarro1999},
\begin{equation}
    B(\mathbf{k}_{1}, \mathbf{k}_{2}) = \sum_{\ell=0}^{\infty} \sum_{m=-\ell}^{\ell} B_{m}^{\ell}(k_{1}, k_{2}, k_{3}) Y_{\ell}^{m}(\mu_{1}, \mu_{2}).
\end{equation}

The first multipole of the bispectrum (monopole) corresponds to $\ell= 0 $ and $ m = 0$, with $Y^0_0=1$. Thus, it can be obtained from the redshift space bispectrum following
\begin{equation}
    B_{g}^{(0)}(k_{1}, k_{2}, k_{3}) = \int d\mu_{1} d\mu_{2} B_{g}(\mathbf{k}_{1}, \mathbf{k}_{2}) \equiv \int_{-1}^{+1} d\mu_{1} \int_{0}^{2\pi} d\varphi \, B_{g}(\mathbf{k}_{1}, \mathbf{k}_{2}),
\end{equation}

where $\varphi$ has been defined to be $\mu_{2} \equiv \mu_{1}x_{12} - \sqrt{1 - \mu_{1}^{2}} \sqrt{1 - x_{12}^{2}} \cos \varphi$, and $x_{12} \equiv (\mathbf{k}_{1} \cdot \mathbf{k}_{2}) / (k_{1} k_{2})$.
Integrating over the line of sight of the two vectors we obtain an expression for the monopole,
\begin{equation}
    B_{g}^{(0)}(\mathbf{k}_{1}, \mathbf{k}_{2}) = \int d\mu_{1} d\mu_{2} B_{g}(\mathbf{k}_{1}, \mathbf{k}_{2}).
\end{equation}
When $D_{\text{FoG}}^{B}=1$, an analytical expression for $B_{g}^{(0)}$ can be obtained.

\begin{figure}[ht]
    \centering
    \includegraphics[width=\textwidth]{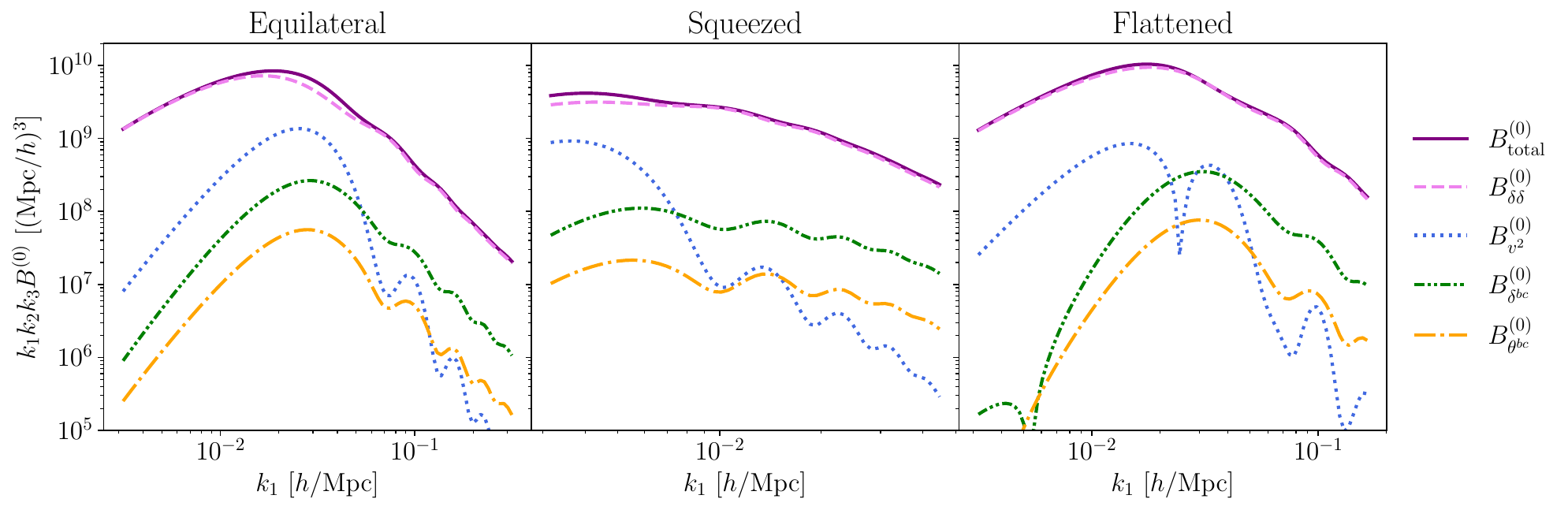}
    \caption{Contributions to the bispectrum for three different triangle configurations, equilateral (left), squeezed (center) and flattened (right). We show the total bispectrum monopole (purple, solid), the nonlinear matter density bispectrum (violet, dashed), and the three relative velocity contributions, weighted by the bias parameters $b_{v^2}=0.01$ (blue, dotted), $b_\delta^{bc}=1$ (green, dash-dot-dotted) and $b_\theta^{bc}=100$ (orange, dash-dotted).}
    \label{fig:contributions_bk}
\end{figure}
Figure \ref{fig:contributions_bk} shows the contributions to the bispectrum monopole for three different triangle configurations, equilateral, squeezed and flattened. For illustrative purposes, we adopt fiducial bias parameter values of $b_{v^2}=0.01$, $ b_\delta^{bc}=1$ and $ b_\theta^{bc}=100$. In the equilateral configuration ($k_1=k_2=k_3$, $\cos\theta_{1,2}    =-0.5$) the three relative velocity contributions peak at $k\sim 0.03h/{\rm Mpc}$, and decreasing sharply at both larger and smaller scales. The $B^{(0)}_{v^{2}}$, $B^{(0)}_{\delta^{bc}}$ and $B^{(0)}_{\theta^{bc}}$ terms exhibit oscillatory features on the same scales as the BAO, analogous to the behavior observed in the power spectrum, directly interfering with the density BAO feature. For the isosceles squeezed configuration ($k_2=k_3,\; k_1/k_2=0.14,\; \cos\theta_{2,3}=-0.99$), the velocity contribution peaks at a smaller $k\sim 0.004\,h/{\rm Mpc}$, corresponding to larger scales. Here, the relative density and divergence contributions remain almost constant in amplitude while retaining their oscillatory feature. For isosceles flattened configurations ($k_1=k_2, \,k_3\sim k_1+k_2,\, \cos\theta_{2,3}=-0.99$) the overall shape of the relative velocity contributions changes. However, a critical universal feature persists across all configurations: the relative velocity terms introduce phase-shifted oscillations that couple to the matter density bispectrum exactly at BAO scales. This interference distorts the overall shape of the bispectrum, creating a potential systematic bias in BAO measurements, if ignored or unmodeled.

\subsection{\texorpdfstring{Setting the scene: relative velocity effect on the BAO of  $P$ and  $B$}{Setting the scene: relative velocity effect on the BAO of  P and  B}}
\label{sec:sett_scene}
Before proceeding with the analysis, we first develop physical intuition by visualizing how the BAO signature in the power spectrum multipoles and bispectrum monopole is modified by the various velocity bias parameters
Starting from a fiducial model we construct ideal synthetic noiseless "measurements" (i.e. using the theory model of sec.~\ref{Sec:RELVEL} and app.~\ref{sec:appendix_A}) for the power spectrum and bispectrum monopole for different values of $b_{v^2}$, $b_\delta^{bc}$ and $b_\theta^{bc}$.
Here and hereafter we use the fiducial cosmology of \texttt{Planck18}$ \Lambda$CDM \cite{Planck2018}, with key parameters $h = 0.6736$, $\omega_{\rm CDM} = 0.1200$, $\omega_b = 0.02237$, $\sigma_8 = 0.811355$, $n_s = 0.9649$, $N_{ur} = 2.0328$ and one massive neutrino with $\omega_\nu = 0.00064420$.

\begin{figure}[ht]
    \centering
    \includegraphics[width=\textwidth]{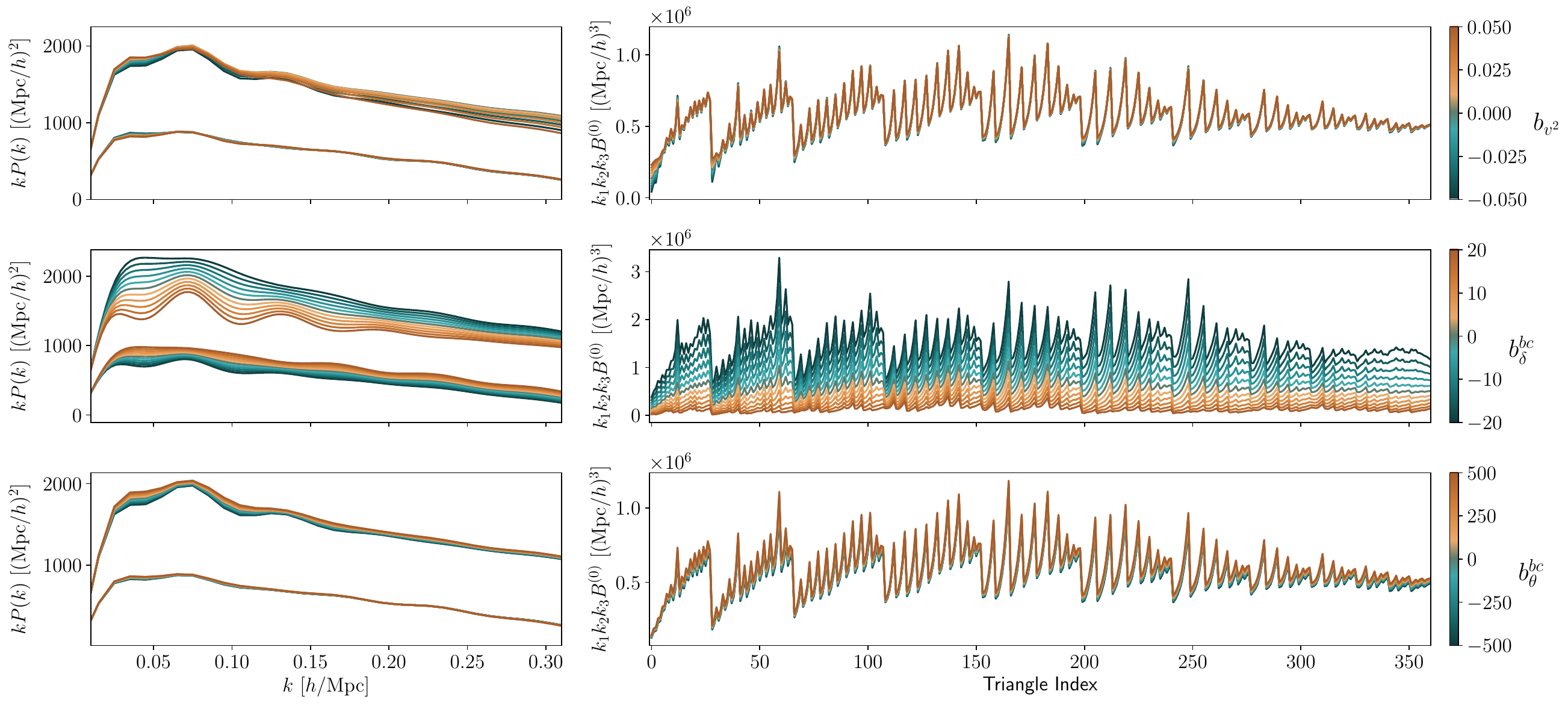}
    \caption{Impact of the relative velocity bias parameters on the large-scale structure observables. We show variations in the power spectrum multipoles (left) and bispectrum monopole (right) data vectors for the range of the relative velocity bias parameters $b_{v^2}$ (top), $b_\delta^{bc}$ (middle) and $b_\theta^{bc}$ (bottom) studied in this work.}
    \label{fig:pkbk_datavectors}
\end{figure}

Figure \ref{fig:pkbk_datavectors} shows the change in the power spectrum and bispectrum data vectors when varying $b_{v^2}$ (top), $b_\delta^{bc}$ (middle) and $b_\theta^{bc}$ (bottom). 
The triangle configurations in the $x$-axis are sorted hierarchically by ascending value of $k_3$, then $k_2$ and finally $k_1$, enforcing $k_1\le k_2\le k_3$ and spanning $0.02\le k_i\,(h/{\rm Mpc})\le0.2$. Under this scheme, the left side of the plot (lower indices) represents configurations probing the largest cosmic scales. 

We observe that the $b_{v^2}$ contribution significantly changes the shape of the power spectrum. The effect is a decreasing of power, combined with a suppression (enhancement) of the wiggle pattern for positive (negative) values of $b_{v^2}$.
For the bispectrum, the effect is intensified for lower values of $k_1$, corresponding to large scales in the LSS. 
When it comes to $b_\delta^{bc}$, the relative velocity contribution results in a notable decrease (increase) of both the power spectrum and the bispectrum monopole amplitudes for positive (negative) values. Alternatively, it produces the opposite effect in the power spectrum quadrupole, changing the monopole-to-quadrupole ratio. 
Finally, for $b_\theta^{bc}$, we observe almost no effect in the power spectrum quadrupole, with a slight amplitude contribution in the monopole. For the bispectrum, it results in a slight enhancement (suppression) of power for positive (negative) values.
We perform this study for the parameter ranges considered in \cite{Beutler2017}, based on the estimations of \cite{Schmidt2016}.

\section{BAO analysis}
\label{sec:BAO}
To quantify the impact of the relative velocity in the determination of the BAO scale we extract the isotropic dilation parameter from the generated synthetic data vectors described in sec.~\ref{sec:sett_scene} and shown in Figure.~\ref{fig:pkbk_datavectors}, under the assumption of no velocity bias, and evaluate how the relative velocity effect biases the measurement. 
To do so, the ingredients needed are the BAO model for the power spectrum (sec.~\ref{sec:BAOPk}) and bispectrum monopole (sec.~\ref{sec:baobisp}) to use in a Gaussian likelihood with the covariance described in sec.~\ref{sec:covariance}.

We begin by presenting the adopted BAO modeling and analysis for the power spectrum and bispectrum. The power spectrum treatment is standard and reported here only for defining the notation adopted. The novel part is the BAO modeling in the bispectrum.

As standard practice, the goal of the BAO analysis is to isolate the BAO signal from the full clustering measurement. This signal is compressed in the parallel and transverse dilation parameters, defined as 
\begin{equation}
    \label{eq:def_alphes}
   \alpha_{\parallel} (z)=\frac{D_H(z)r_{\text{drag}}^{\text{ref}}}{D_H^{\text{ref}}(z)r_{\text{drag}}}, \quad\quad  \alpha_{\perp}(z)=\frac{D_M(z)r_{\text{drag}}^{\text{ref}}}{D_M^{\text{ref}}(z)r_{\text{drag}}},
\end{equation}
where $D_H(z)=c/H(z)$, $H(z)$ is the Hubble expansion parameter, $r_{\text{drag}}$ the commoving sound horizon at drag epoch, and $D_M(z)$ is the commoving angular diameter distance at redshift $z$. The superscript "ref" denotes quantities measured in a reference cosmology. A common reparameterization of these parameters gives the geometric mean, $\alpha_{\rm iso}$, and the ratio across and along the line of sight of the dilation, $\alpha_{\rm AP}$, defined as:
\begin{equation}
    \alpha_{\text{iso}}=(\alpha_{\parallel}\alpha_{\perp}^{2})^{1/3}, \quad \quad \alpha_{\rm AP}=\alpha_{\parallel}/\alpha_{\perp}.
\end{equation}
The standard methodology adopted in major spectroscopic surveys \cite{eBossBAO,DESIDR1} is a template-based measurement of the dilation parameters employing a parameterized model to marginalize over non-linearities present in the broadband signal. Because this method cannot isolate the standard BAO from other oscillatory signatures within the power spectrum, the measurement remains vulnerable to systematic biases introduced by the relative velocity effect and analogous phase-shifted signals. This potential bias has been quantified previously for the power spectrum BAO \cite{Beutler2014}, and the bispectrum has been shown to be an important tool to disentangle these oscillatory features \cite{Yoo2011}. 

In this work we present a new methodology to extend the BAO measurement to the bispectrum monopole  and evaluate its ability to differentiate the relative velocity effect within the BAO signal. While both the standard power spectrum and our new bispectrum BAO extraction techniques are blind to this phase-shifted signal, the measurement from the bispectrum can serve as a tool to identify and estimate the presence of the relative velocity contribution, thereby mitigating systematic bias.
We next describe the fitting models used.

\subsection{BAO model in the Power Spectrum}
\label{sec:BAOPk}
The BAO model for the power spectrum we employ aims to capture the anisotropic BAO information present in the multipoles, following the analysis from \cite{eBossBAO}.
The Alcock-Paczynski effect \cite{AP_effect} and the isotropic dilation distort the wavenumbers along and across the line of sight, giving the observed wavenumbers $k_\parallel^{\rm obs}$ and $k_\perp^{\rm obs}$, which are related to the true ones, $k_\parallel'$ and $k_\perp'$, following $k_\parallel'=k_\parallel^{\rm obs}\alpha_\parallel$ and $k_\perp'=k_\perp^{\rm obs}\alpha_\perp$. In terms of $\alpha_{\rm iso}$ and $\alpha_{\rm AP}$, the absolute wavenumber, $k=\sqrt{k_\parallel^2+k_\perp^2}$, and the cosine of the angle between the wavenumber vector and the line-of-sight (LOS) direction, $\mu$, 

\begin{equation}
k' = \frac{\alpha_{\mathrm{AP}}^{1/3}}{\alpha_{\mathrm{iso}}}
\left[ 1 + \mu_{\mathrm{obs}}^2 
\left( \frac{1}{\alpha_{\mathrm{AP}}^2} - 1 \right) \right]^{1/2}
k_{\mathrm{obs}}, \quad \quad
\mu' = \frac{\mu_{\mathrm{obs}}}{\alpha_{\mathrm{AP}}}
\left[ 1 + \mu_{\mathrm{obs}}^2 
\left( \frac{1}{\alpha_{\mathrm{AP}}^2} - 1 \right) \right]^{-1/2}.
\end{equation}

We model the anisotropic power spectrum following
\begin{equation}
    P(k,\mu)=B_p(1+R\beta\mu^2)^2P^{(\rm sm)}_{\text{lin}}(k)\left\{1+[\mathcal{O}_{\text{lin}}(k)-1] \times e^{-k^2[\mu^2\Sigma_{\parallel}^2+(1-\mu^2)\Sigma_{\perp}^2]/2}\right\}.
\end{equation}
Here, $P_{\text{lin}}$ is the linear power spectrum and $\mathcal{O}_{\text{lin}}\equiv P_{\text{lin}}/P^{(\text{sm})}_{\text{lin}}$ is the linear BAO template, where $P^{(\text{sm})}_{\text{lin}}$ is a smoothed power spectrum with no BAO signal.
$B_p$ is identified with the linear bias squared, $b_1^2$, $\beta$ is the redshift space distortion parameter, and $R$ accounts for the RSD suppression due to reconstruction. 
For a pre-reconstructed field and when reconstruction preserves RSD, $R=1$. For reconstruction conventions that don't preserve RSD, $R=1-\exp{(-k^2\Sigma_s/2)}$, with $\Sigma_s$ the smoothing scale used during the reconstruction process. 
The damping of the BAO due to bulk motions of galaxies is described by the parameters $\Sigma_{\parallel}$ and $\Sigma_{\perp}$, along and across the LOS, respectively.

The multipoles are obtained by weighting $P(k,\mu)$ by the Legendre polynomials of $\mu$, $\mathcal{L}_\ell(\mu)$, to obtain
\begin{equation}
    P^{(\ell)}(k) =\frac{2\ell + 1}{2}\int_{-1}^{1} d\mu \,\mathcal{L}_\ell(\mu)\,P\!\left[k'(k,\mu),\, \mu'(\mu)\right]+ \sum_{i=1}^{n} A_{p,i}^{(\ell)} k^{2-i}.
\end{equation}
The $A_i$ coefficients allow us to marginalize over non-linear effects present in the broadband.

We fit $\alpha_{\parallel}$ and $\alpha_{\perp}$, as well as $B$, $\beta$, $\Sigma_\parallel$, $\Sigma_\perp$ and 5 coefficients per multipole for the polynomial describing the broadband of the power spectrum. We do so by exploring the likelihood surface using the MCMC solver \textsc{cobaya}\footnote{\href{https://github.com/CobayaSampler/cobaya}{https://github.com/CobayaSampler/cobaya}}, with a Gelman-Rubin convergence criterion of $R_{\rm GR}-1\leq0.01$.
We perform the fit in the range $0.02\le k\,(h/{\rm Mpc})\le0.3$.
The measurements of $\alpha_{\parallel}$ and $\alpha_{\perp}$ are always marginalized over the rest of the nuisance parameters, including $B$, $\beta$, $\Sigma_{\parallel,\perp}$ and the 5 coefficients per multipole. With this analysis, we can capture the BAO anisotropic information in the power spectrum multipoles. 

\subsection{Isotropic BAO model in the Bispectrum}
\label{sec:baobisp}
The \CH{}bispectrum has been shown to contain information that complements that of the power spectrum \cite{Samushia2017}. In \cite{Samushia2018} the BAO feature is detected for the first time in the bispectrum. In \cite{HChild2018a} and \cite{HChild2018b}, its interferometric nature is explored to quantify the constraining improvement of the scaling parameters when taking the bispectrum into account. Additionally, in \cite{Behera2024} the isotropic dilation parameter, $\alpha_{\rm iso}$, was successfully extracted from the bispectrum monopole, but without an estimation of the constraining power of the method due to the lack of a reliable covariance. We aim to model the signal and produce an estimation of both its accuracy and precision for measuring $\alpha_{\rm iso}$ from the bispectrum monopole and its performance when applied to a joint power spectrum and bispectrum analysis. 
We focus our efforts on the monopole component of the bispectrum, since measuring higher-order multipoles of the redshift-space galaxy bispectrum is still challenging and computationally expensive, and the monopole exhibits a significantly higher signal-to-noise ratio. Likewise, the high dimensionality of the data vector, as well as the construction of the covariance, make it very impractical, at the present time, to use more than a fraction of the available bispectrum measurements for the analysis.

To capture the BAO signal in the bispectrum we propose a similar approach to that for the power spectrum, based on the tree-level bispectrum monopole in redshift space \cite{Scoccimarro1999},
\begin{align}
\label{eq:bispectrumtree}
    B^{(0)}(k_1,k_2,k_3;\alpha_{\rm iso})&=P(k_1)P(k_2)[ \mathcal{M}_1(\beta_F;k_1,k_2,k_3)F_2^{\rm eff}(k_1,k_2,k_3) \nonumber\\& + \mathcal{M}_2(\beta_G;k_1,k_2,k_3)G_2^{\rm eff}(k_1,k_2,k_3) \nonumber\\&+\frac{1}{2}\mathcal{M}_1(\beta_F;k_1,k_2,k_3)(C_1S_2(k_1,k_2,k_3)+C_2)\nonumber\\&+ \mathcal{M}_3(\beta_\mu;k_1,k_2,k_3) ] +{\rm cyc.}
\end{align}
where $\mathcal{M}_i$ are functions of single free parameters, $\beta_i$, which account for the bias and the effect of redshift space distortions of the broadband. We use the  $F^{\rm eff}$ and $G^{\rm eff}$ effective kernels proposed in \cite{hgm2012b,hgm2014b}, and $S_2$ is the non-local bias kernel. The expressions for $\mathcal{M}_i$ and the kernels can be found in Appendix \ref{sec:appendix_B}. In particular, for a dark matter field in real space we expect to have $\mathcal{M}_1=1$ and $\mathcal{M}_2=\mathcal{M}_3=0$.
$P(k)$ stands for the dark matter non-linear power spectrum in real space, which we propose to expand {\it à la} BAO following the methodology used in the power spectrum. 
In eq.~\ref{eq:bispectrumtree} the isotropic signal of the bispectrum is captured following the template introduced in \cite{Anderson2014,HGM2016b},
\begin{equation}
    P(k;\alpha_{\rm iso})=\left[B_bP^{(\rm sm)}_{\rm{lin}}(k)+\sum^n_{i=1}A_{b,i}^{(0)}k^{2-i}\right]\times\left\{1+[\mathcal{O}_{\rm{lin}}(k/\alpha_{\rm iso})-1] e^{-\frac{1}{2}k^2(\Sigma^B_{\rm nl} )^2}\right\}.
\end{equation}
\CH{With this approach, we aim to capture the isotropic dilation parameter $\alpha_{\rm iso}$ by fitting the monopole of the bispectrum. This template doesn't include an explicit $\mu$-dependence like the anisotropic one used for the power spectrum through the damping terms along and across the line of sight $\Sigma_{\parallel}$, $\Sigma_{\perp}$. Therefore, we do not integrate over the line-of-sight, but use an effective weighted-averaged damping parameter, $\Sigma^B_{\rm nl}$. The anisotropic modeling of the damping becomes important for extracting the BAO in the anisotropic signal of clustering, which is left for future work.}
We marginalize over $B_b$, $\Sigma^B_{\rm nl}$, $\beta_F$, $\beta_G$, $\beta_{\mu}$, $C_1$, $C_2$ and 5 coefficients $A_{b,i}^{(0)}$ to absorb the broadband of the bispectrum monopole. 

Having this model, we can fit the bispectrum monopole measurement independently and obtain a measurement for $\alpha_{\rm iso}$ only. Additionally, we can fit the power spectrum and bispectrum jointly, constraining $\alpha_{\rm iso}$ simultaneously, while all other parameters are fit independently for each observable.

Given the novelty of this method, we present a comprehensive validation of the technique.
The performance of this bispectrum model is quantified against N-body simulations in appendix~\ref{sec:appendix_B}. We establish its robustness for BAO extraction in appendix~\ref{sec:appendix_C}. 
We find that this model of the bispectrum performs well up to $k\,(h/{\rm Mpc})\sim0.3$, capturing the shape of the measurement within 5$\sigma$ for a volume of 500($h^{-1}$Gpc)$^3$, except for some flattened configurations. 
We successfully isolate the isotropic BAO component in the bispectrum monopole signal, obtaining an unbiased measurement of $\alpha_{\rm iso}$ with constraining power comparable to the pre-reconstruction power spectrum analysis. Furthermore, a joint power spectrum and bispectrum fit tightens the constraints on $\alpha_{\rm iso}$ by $\sim30\%$.

\subsection{Covariance}
\label{sec:covariance}
Our analysis is done using mock-based covariances. Due to the large size of the bispectrum data vector employed we require a large number of mock realizations to compute our covariance. We use the \textsc{Quijote} suite\footnote{\href{https://quijote-simulations.readthedocs.io/}{https://quijote-simulations.readthedocs.io/}} \cite{Quijote}, consisting of 15000 independent realizations of a N-body simulations of a cubic box of 1($h^{-1}$Gpc)$^3$ volume with periodic boundary conditions, and containing $512^3$ cold dark matter particles. 
We use the dark matter halo catalogs of the $z=0.5$ snapshot, matching the effective redshift of the DESI LRG1 tracers. \CH{Due to the nature of dark matter halos, the FoG effects are typically reduced compared to galaxy catalogs, although the magnitude of the effect depends on the specific tracer.}
Dark matter halos are identified using the Friends-of-Friends (FoF) algorithm, and only saved when containing at least 20 cold dark matter particles, which makes the halo mass cut $M\sim 2\times10^{13} h^{-1}M_{\odot}$. The resulting mean halo number density is $\Bar{n}\sim 5.1 \times 10^{-5} (h/{\rm Mpc})^{-3}$. 
\CH{The measurements of the simulated catalogs are highly consistent with our synthetic noiseless data vectors at large scales, making them well suited to estimate the covariance. The mean measurement of the 15000 realizations presents an amplitude difference that is lower than $\sim10\%$ up to $k=0.3$ ($k=0.15$) for the power spectrum monopole (quadrupole) and $k=0.1$ for the bispectrum monopole. At higher values of $k$, corresponding to smaller physical scales, non-linearities become more important, and we observe a slightly larger discrepancy arising from our theoretical model's limitations in capturing them.}
\CH{In this work we use a numerical covariance to describe the full power spectrum and bispectrum data vector, including their cross-covariance. It is computed using 15000 mock realizations and employed on a size 417 data vector, leading to a Hartlap correction \cite{Hartlap2007} of $\sim0.97$}. 

\section{Results}
\label{sec:Results}
\CH{Performing a full parameter inference is not the objective of this work. Here we focus on the role of the BAO feature in the bispectrum in mitigating biases introduced by velocity bias in the BAO of the  power spectrum.  Hence we have proceeded as follows.}
We have fitted the theoretical data vectors computed in sec.~\ref{Sec:RELVEL}, \CH{with varying relative velocity parameters  $b_{v^2}$, $b_{\delta^{bc}}$ and $b_{\theta^{bc}}$ individually.} We have employed the BAO pipeline described in sec.~\ref{sec:BAO} on different data vector combinations: the power spectrum monopole and quadrupole only, the bispectrum monopole only and a joint fit considering the two.
Since the data is theoretical and noiseless, any deviation from $\alpha_{\rm iso}=1$ corresponds to a bias in the method, introduced by the relative velocity effect.
\CH{The covariance assumed for the analysis corresponds to the statistical error of a survey volume of 500($h^{-1}$Gpc)$^3$, corresponding to a volume of a Stage V surveys and about 10 times that of the complete DESI survey \cite{Schlegel2022,Schlegel2022b}.}

\begin{figure}[ht]
    \centering
    \includegraphics[width=\textwidth]{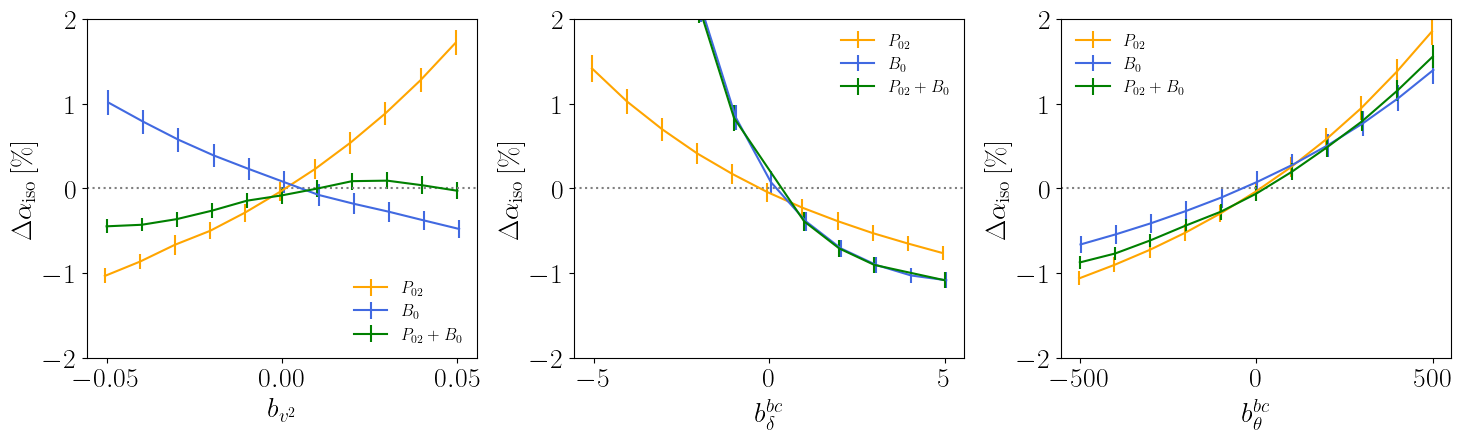}
    \caption{Shift in the $\alpha_{\rm iso}$ parameter as a function of the studied relative velocity bias parameters for the power spectrum only ($P_{02}$, orange), bispectrum only ($B_0$, blue) and joint (Joint $P_{02} + B_0$, green) fits. Left panel shows the shift for varying values of $b_{v^2}$, central panel for $b_{\delta^{bc}}$ and right panel for $b_{\theta^{bc}}$. The displayed error bars correspond to the statistical error from a volume of 500($h^{-1}$Gpc)$^3$.}
    \label{fig:iso_shift}
\end{figure}

Figure \ref{fig:iso_shift} shows the shift in $\alpha_{\rm iso}$ recovered, $\Delta\alpha_{\rm iso}=\alpha^{\rm measured}-\alpha^{\rm true}$, depending on the amplitude of the three relative velocity parameters. We show the measured shift from the expected value when fitting the power spectrum multipoles only ($P_{02}$, orange), the bispectrum monopole only ($B_0$, blue), and fitting them in a joint analysis ($P_{02} + B_0$, green).  
We observe three different behaviors for each bias contribution.
For the $b_{v^2}$ parameter we obtain the best-case scenario, in which the contributions to the power spectrum and the bispectrum shift $\alpha_{\rm iso}$ in opposite directions. We observe that for the $P_{02}$ fit, the recovered shift evolves from $-1\%$ when $b_{v^2}=-0.05$ to $\sim2\%$ when $b_{v^2}=0.05$. For the $B_0$ only fit, the behavior shows the opposite trend, with the shift ranging from $1\%$ to $-1\%$ for the same range. The joint $P_{02} + B_0$ fit lies in the middle, with values of $\alpha_{\rm iso}$ that try to compensate both behaviors.
Therefore, with the two independent measurements of $\alpha_{\rm iso}$, from $P_{02}$ and $B_0$, this effect can be identified and constrained.

For $b_{\delta^{bc}}$, we find the initially explored range ($-20\le b_{\delta^{bc}}\le20$) yields unphysical results, inducing highly non-perturbative distortions in the measurement's broadband amplitude and shape. The resulting BAO fits exhibit extreme systematic shifts in the isotropic dilation parameter, with $\alpha_{\rm iso}$ shifting by up to $20\%$ when fitting the bispectrum for values of $b_\delta^{bc}\le-8$. The model can no longer successfully capture the bispectrum shape; therefore, we restrict our parameter prior. Specifically, we confine our analysis to the $95\%$ confidence interval reported in recent literature \cite{Beutler2017}.
The power spectrum fits show a tendency in the shift of $\alpha_{\rm iso}$ that evolves monotonically from $1.5\%$ when $b_{\delta^{bc}}=-5$ to $\sim-0.5\%$ when $b_{\delta^{bc}}=5$.
The $B_{\delta^{bc}}^{(0)}$ contribution to the bispectrum rapidly degrades the fit of our model and the recovered $\alpha_{\rm iso}$ for $b_\delta^{bc}<-2$, reaching a shift of $10\%$ on $\alpha_{\rm iso}$ for $b_\delta^{bc}<-5$. For positive values of the bias parameter it follows a similar trend to that of the power spectrum, peaking at $-1\%$. 
The joint $P_{02} + B_0$ fit follows closely the bispectrum fits tendency, showing how the $B_{\delta^{bc}}^{(0)}$ contributions dominate this effect.

Finally, for $b_{\theta^{bc}}$, the power spectrum and bispectrum contributions of the relative velocity effect result in a similar shift for $\alpha_{\rm iso}$. The tendency of all three fits follows the one observed in the $P_{02}$ fit for the $b_{v^2}$ parameter. In this scenario, all three fits evolve coherently from $-1\%$ to $\sim2\%$ for the studied range of $b_{\theta^{bc}}$. For this parameter, the bispectrum shows a slightly smaller response to the relative velocity contribution. Therefore, the $B_0$ fit has a limited capacity to disentangle the effect of $b_{\theta^{bc}}$ from the $P_{02}$ fit.

Previous measurements find $b_{r}<0.33$ \cite{Yoo2013}, which translates to $b_{v^2}<0.1$ at $95\%$ confidence level using the power spectrum of the SDSS-III Data Release 9 CMASS sample \cite{SDSSDR9,CMASS,BOSSpkDR9}. 
Observational evidence pointing to significant values of $b_{\delta}^{bc}$ was presented in \cite{Soumagnac2016}.
In \cite{Slepian2018}, the 3-point correlation function of the BOSS DR12 CMASS sample \cite{BOSS3PCFDR12} is used to find $b_{v^2}<0.01$. More recently, \cite{Beutler2017} found $b_{v^2}=0.012 \pm 0.015 (\pm 0.031)$, $b_{\delta^{bc}}=-1.0 \pm 2.5 (\pm6.2)$ and $b_{\theta^{bc}}=-114 \pm 55 (\pm175)$ with $68\%$ ($95\%$) confidence interval.
Taking the latest measurement, our results show that the bispectrum BAO is highly sensitive to the distinct signatures of the three bias parameters within the $95\%$ confidence interval.
Since the $b_{\delta^{bc}}$ and $b_{v^2}$ terms affect $P_{02}$ and $B_0$ differently, the complementarity between these observables can be used to detect this effect or to mitigate its impact.
Furthermore, a joint $P_{02} + B_0$ fit will give a better constrain on $\alpha_{\rm iso}$, as shown in Appendix \ref{sec:appendix_B}.

\begin{figure}[ht]
    \centering
    \includegraphics[width=\textwidth]{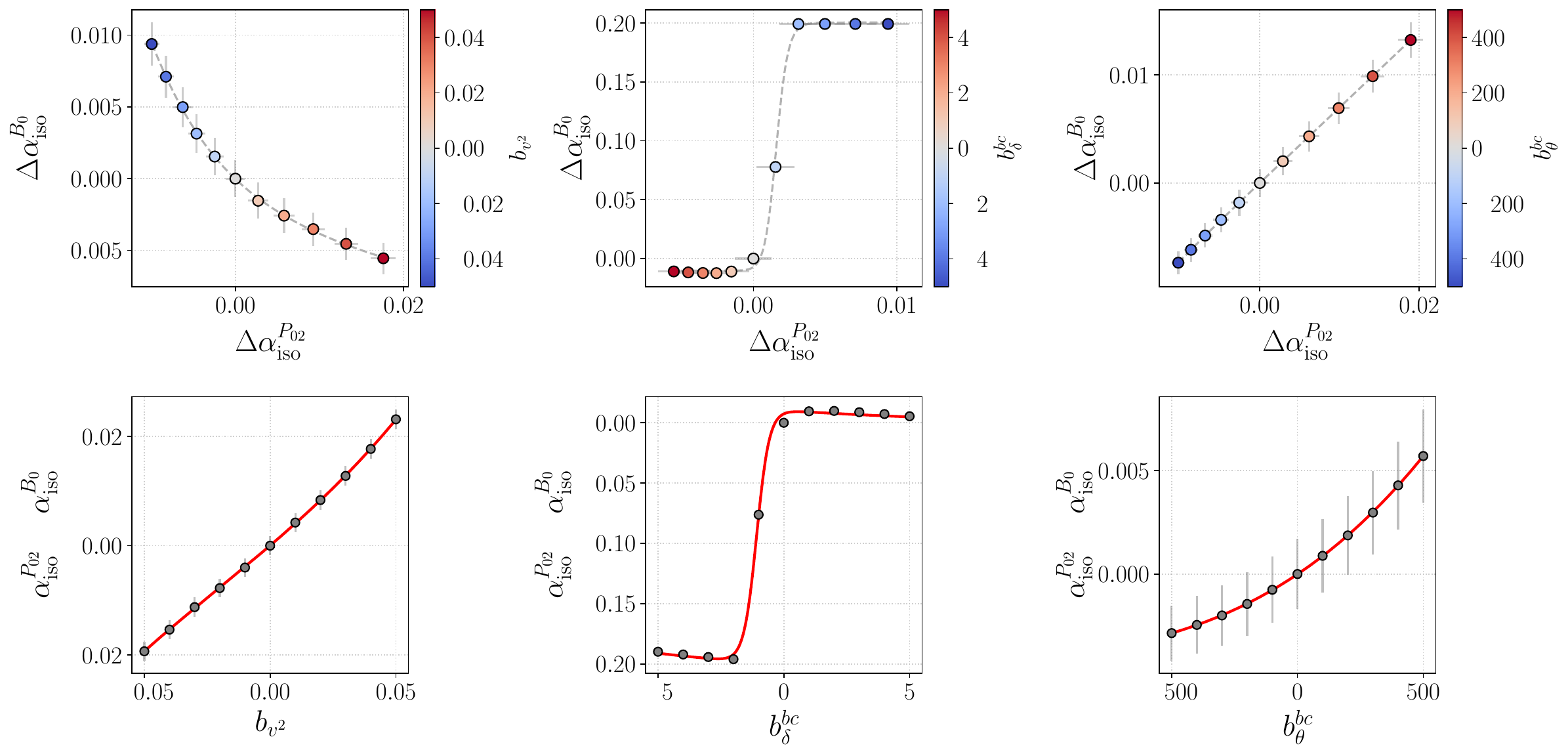}
    \caption{Isotropic dilation parameter measurements as a function of the relative velocity bias parameters. Top row shows how the recovered isotropic dilation parameter shift ($\Delta\alpha_{\rm iso}$), derived independently from the power spectrum and the bispectrum monopole, evolves for each of the three bias parameters. Bottom row shows the disparity between the power spectrum and bispectrum measurements across the same parameter ranges. The error bars displayed in all plots correspond to the statistical error from a volume of 500($h^{-1}$Gpc)$^3$.}
    \label{fig:inverse_shift}
\end{figure}
Figure~\ref{fig:inverse_shift} further illustrates how two independent measurements of $\alpha_{\rm iso}$, from $P_{02}$ and from $B_0$, can serve as a tool to detect and disentangle the velocity bias parameters' respective impacts on large-scale galaxy clustering.
The top row shows the dependence of the recovered $\Delta\alpha_{\rm iso}^{P_{02}}$ and $\Delta\alpha_{\rm iso}^{B_0}$, as a function of the amplitude of the relative velocity bias parameter. The displayed error bars correspond to a volume of 500($h^{-1}$Gpc)$^3$. 
We can see the different response of the recovered BAO scale from $P$ and $B$ for the same change in bias parameter.
Especially in the central panel, we observe how the $\alpha_{\rm iso}$ measured from the bispectrum rapidly diverges for $b_{\delta^{bc}}<0$, pushing against the edge of the imposed prior on $\alpha_{\rm iso}^{B_0}$. Due to the scale of the plot, the error bars associated with this measurement are not visible. The top panels are only illustrative: the BAO technique is blind to the measurement of the true $\Delta\alpha_{\rm iso}$, because that would require knowing the true cosmology of the universe, so the relevant quantity to study these parameters is the difference between the power spectrum and bispectrum measurements.
The bottom row shows the difference between these measurements as a function of the amplitude of each relative velocity contribution. We fit the points to a polynomial fit (for $b_{v^2}$ and $b_{\theta^{bc}}$) or a sigmoid (for $b_{\delta^{bc}}$). We observe sensitivity to $b_{v^2}$, and drastic response for $b_{\delta^{bc}}<0$, demonstrating promising potential in placing competitive constraints on these effects. For a DESI-like survey, with a volume of $\sim 50(h^{-1}$Gpc)$^3$, the difference between both measurements would hide within the statistical error for the explored range of $b_{\theta^{bc}}$.

\section{Conclusions}
\label{Sec:Conclusions}

We have modeled the relative velocity effect in galaxy clustering, both the power spectrum monopole and quadrupole and bispectrum monopole in redshift space and tested its effect on the BAO measurement. 
The novel results of this work are:
\begin{itemize}
    \item We have extended the relative velocity framework for galaxy clustering by presenting an updated model for the redshift space tree-level bispectrum to include all relative velocity terms.
    \item We present a novel technique to measure the isotropic dilation parameter from the BAO signal in the bispectrum monopole. We have validated the performance of the model on N-body simulations and tested its ability to extract $\alpha_{\rm iso}$. Our study finds it is an unbiased technique that successfully recovers the isotropic dilation scale. Furthermore, when combined with the power spectrum measurement, it is able to increase by $~30\%$ the constraining power of the pre-reconstruction analysis, complementing the post-reconstruction standard technique. 
    \item We have investigated how each relative velocity contribution modifies the power spectrum and bispectrum measurements. We have tested how these introduce a bias in the recovery of $\alpha_{\rm iso}$. Our results show that each relative velocity contribution introduces a systematic shift that alters differently the recovered $\alpha_{\rm iso}$ from the power spectrum and bispectrum. We find a $2\%$ difference between both measurements for $b_{v^2}=\pm 0.05$, and up to a $20\%$ difference for $b_{\delta^{bv}}\le-2$, suggesting we can estimate the amplitude of these bias parameters from comparing the two independent BAO measurements.
    \item We provide a prescription to identify the effect of each relative velocity contribution and constrain the three associated bias parameters using the BAO in the bispectrum, in combination with the standard power spectrum measurement. We find this method to be optimal for $b_{v^2}$, and highly sensitive to negative values of $b_{\delta^{bc}}$.
\end{itemize}

Standard BAO analyses remain vulnerable to several systematic effects. The relative velocity between baryons and cold dark matter is known to be a potential source of bias for the well-established BAO measurement in the power spectrum, eventually biasing the cosmological interpretation if undetected. 
The bispectrum BAO is not free of this bias, however, due to the different behavior this systematic exhibits, a discrepancy between the $\alpha_{\rm iso}^{P_{02}}$ and $\alpha_{\rm iso}^{B_{0}}$ measurements can help constrain the relative velocity bias parameters.

\CH{We find that for a survey volume equivalent to the full DESI footprint, our methodology can successfully detect velocity biases of $|b_{v^2}|\ge0.02$ and $b_{\delta^{bv}}\le0.5$. However, the current approach lacks sensitivity to the effects introduced by the $b_{\theta^{bv}}$ terms, a limitation that persists even when considering a volume ten times larger than DESI. Future work extending this analysis to incorporate the full shape of the bispectrum could further constrain these remaining effects.}

Our work establishes the BAO in the bispectrum as a cosmological probe that can be measured in current and future surveys. It can not only serve as a complementary measurement to test the standard analysis, but also as a tool to detect and constrain the relative velocity effect, as well as other proposed models with analogous features.
\acknowledgments

ANM and HGM acknowledge support through the Consolidación Investigadora (CNS2023-144605) of the Spanish Ministry of Science and Innovation. The authors acknowledge support of Proyectos de Generación de Conocimiento (PID2022-141125NB-I00) of the Spanish Ministry of Science and Innovation and the “Center of Excellence Maria de Maeztu 2025-2029” award to the ICCUB funded by grant CEX2024-001451-M from AEI/10.13039/501100011033.

\appendix
\section{Power Spectrum Model Terms}
\label{sec:appendix_A}
For our redshift space power spectrum model, we follow the approach of \cite{Beutler2017}, which builds on the Eulerian non-linear and non-local bias model proposed in \cite{McDonaldRoy2009}. The full expression including all relative velocity terms reads
\begin{align}
P_g(k,\mu)&= P_{g,\mathrm{NL}}(k,\mu)+ b_{v^2}\Big[b_1 P_{\delta|v^2}(k)+ b_2 P_{\delta^2|v^2}(k)+ b_{s^2} P_{s^2|v^2}(k)+ b_{v^2} P_{v^2|v^2}(k)\Big]
\nonumber\\&\quad
+ b_1 b_{v^2} P_{\mathrm{adv}|\delta}(k)+ 2 b_1 b_\delta^{\mathrm{bc}} P_{\delta|\delta_{\mathrm{bc}}}(k)+ 2 b_1 b_\theta^{\mathrm{bc}} P_{\delta|\theta_{\mathrm{bc}}}(k) \nonumber\\&\quad
- 2 f \mu^2\Big[b_{v^2}\big(b_1 P_{\delta|v^2 v_\parallel}(k)+ P_{\mathrm{adv}|v_\parallel}(k)\big)- b_\theta^{\mathrm{bc}} P_{\delta|\theta_{\mathrm{bc}}}(k)+ b_\delta^{\mathrm{bc}} P_{\delta|\delta_{\mathrm{bc}}}(k) \nonumber\\&\qquad\qquad
+ b_{v^2}\big(P_{v^2|v_\parallel}(k)+ P_{v^2|\delta v_\parallel}(k)\big)\Big]\nonumber\\&\quad+ f^2 \mu^4\, b_{v^2}\,P_{v_\parallel|v^2 v_\parallel}(k)- f^2 \mu^2\, b_{v^2}\left[I_1(k) + \mu^2 I_2(k)\right].
\end{align}

Under the assumption that galaxy bias is local in Lagrangian space \cite{Chan2012,Baldauf2012,Saito2014},
\begin{align}
b_{s^2} &= -\frac{4}{7}(b_1 - 1), \\
b_{3n1} &= \frac{32}{315}(b_1 - 1).
\end{align}

The redshift space galaxy non-linear power spectrum, $P_{g,\mathrm{NL}}$, is given by \cite{McDonaldRoy2009,Beutler2014}
\begin{align}
P_{g, \rm NL}(k, \mu) &= D_{\mathrm{FoG}}^P\left(k, \mu, \sigma_{\mathrm{FoG}}^P\right) \Big[ 
P_{g,\delta\delta}(k) + 2f\mu^2 P_{g,\delta\theta}(k) + f^2\mu^4 P_{\theta\theta}(k) \notag \\
&\quad + b_1^3 A^{\rm TNS}(k, \mu, f/b_1) + b_1^4 B^{\rm TNS}(k, \mu, f/b_1) \Big],
\end{align}
with the Fingers-of-God term:
\begin{equation}
    D_{\text{FoG}}^{P}(k, \mu, \sigma_{\text{FoG}}^{P}) = \left( 1 + k^{2}\mu^{2}(\sigma_{\text{FoG}}^P)^{2}/2 \right)^{-2}.
\end{equation}
The $A^{\rm TNS}$ and $B^{\rm TNS}$ terms, introduced in \cite{TNS2010,TN2011}, arise to correct the coupling between the Kaiser and  Fingers-of-God effects. Their expressions are given by 
\begin{align}
    A^{\text{TNS}}(k, \mu, b) &= (k \mu f) \int \frac{d^{3} \mathbf{q}}{(2\pi)^{3}} \frac{q_{z}}{q^{2}} \left\{ B_{\sigma}(\mathbf{q}, \mathbf{k} - \mathbf{q}, -\mathbf{k}) - B_{\sigma}(\mathbf{q}, \mathbf{k}, -\mathbf{k} - \mathbf{q}) \right\}, \\
    B^{\text{TNS}}(k, \mu, b) &= (k \mu f)^{2} \int \frac{d^{3} \mathbf{q}}{(2\pi)^{3}} F^{\text{TNS}}(\mathbf{q}) F^{\text{TNS}}(\mathbf{k} - \mathbf{q}),
\end{align}

where
\begin{equation}
    F^{\text{TNS}}(\mathbf{q}) \equiv \frac{q_{z}}{q^{2}} \left[ b_{1} P_{\delta \theta}(q) + f \frac{q_{z}^{2}}{q^{2}} P_{\theta \theta}(q) \right],
\end{equation}
and
\begin{equation}
    (2\pi)^{3} \delta_{D}(\mathbf{k}_{123}) B_{\sigma}(\mathbf{k}_{1}, \mathbf{k}_{2}, \mathbf{k}_{3}) \equiv \left\langle \theta(\mathbf{k}_{1}) \left\{ b_{1} \delta(\mathbf{k}_{2}) + f \frac{k_{2z}^{2}}{k_{2}^{2}} \theta(\mathbf{k}_{2}) \right\} \left\{ b_{1} \delta(\mathbf{k}_{3}) + f \frac{k_{3z}^{2}}{k_{3}^{2}} \theta(\mathbf{k}_{3}) \right\} \right\rangle.
\end{equation}

The real space galaxy density and divergence auto- and cross-correlations are
\begin{align}
P_{g,\delta\delta}(k) &= b_1^2 P_{\delta\delta}^{\text{SPT}}(k) + 2b_2 b_1 P_{b_2,\delta}(k) + 2b_{s^2} b_1 P_{bs2,\delta}(k) \notag \\
&\quad + b_2^2 P_{b22}(k) + 2b_2 b_{s2} P_{b2s2}(k) + b_{s^2}^2 P_{bs22}(k) + 2b_1 b_{3\mathrm{nl}} \sigma_3^2(k) P^{\mathrm{lin}}(k)\\
P_{g,\delta\theta}(k) &= b_1 P_{\delta\theta}^{\text{SPT}}(k) + b_2 P_{b_2,\theta}(k) + b_{s^2} P_{bs2,\theta}(k)+  b_{3\mathrm{nl}} \sigma_3^2(k) P^{\mathrm{lin}}(k)\\
P_{g,\theta\theta}(k) &=  P_{\theta\theta}^{\text{SPT}}(k)
\end{align}

The integral expressions for the 1-loop redshift space galaxy power spectra terms are \cite{McDonaldRoy2009,Beutler2014},
\begin{align}
P_{b_2,\delta} &= \int \frac{d^3q}{(2\pi)^3} P^{\mathrm{lin}}(q) P^{\mathrm{lin}}(|\mathbf{k} - \mathbf{q}|) {F}_2(\mathbf{q}, \mathbf{k} - \mathbf{q}) \\
P_{bs2,\delta} &= \int \frac{d^3q}{(2\pi)^3} P^{\mathrm{lin}}(q) P^{\mathrm{lin}}(|\mathbf{k} - \mathbf{q}|) {F}_2(\mathbf{q}, \mathbf{k} - \mathbf{q}) S_2(\mathbf{q}, \mathbf{k} - \mathbf{q}) \\
P_{b_2,\theta} &= \int \frac{d^3q}{(2\pi)^3} P^{\mathrm{lin}}(q) P^{\mathrm{lin}}(|\mathbf{k} - \mathbf{q}|) {G}_2(\mathbf{q}, \mathbf{k} - \mathbf{q}) \\
P_{bs2,\theta} &= \int \frac{d^3q}{(2\pi)^3} P^{\mathrm{lin}}(q) P^{\mathrm{lin}}(|\mathbf{k} - \mathbf{q}|) {G}_2(\mathbf{q}, \mathbf{k} - \mathbf{q}) S_2(\mathbf{q}, \mathbf{k} - \mathbf{q}) \\
P_{b2s2} &= -\frac{1}{2} \int \frac{d^3q}{(2\pi)^3} P^{\mathrm{lin}}(q) \left[ \frac{2}{3} P^{\mathrm{lin}}(q) - P^{\mathrm{lin}}(|\mathbf{q} - \mathbf{k}|) S_2(\mathbf{q}, \mathbf{k} - \mathbf{q}) \right] \\
P_{bs22} &= -\frac{1}{2} \int \frac{d^3q}{(2\pi)^3} P^{\mathrm{lin}}(q) \left[ \frac{4}{9} P^{\mathrm{lin}}(q) - P^{\mathrm{lin}}(|\mathbf{k} - \mathbf{q}|) S_2(\mathbf{q}, \mathbf{k} - \mathbf{q})^2 \right] \\
P_{b22} &= -\frac{1}{2} \int \frac{d^3q}{(2\pi)^3} P^{\mathrm{lin}}(q) \left[ P^{\mathrm{lin}}(q) - P^{\mathrm{lin}}(\mathbf{k} - \mathbf{q}) \right] \\
\sigma_3^2(k) &= \int \frac{d^3q}{(2\pi)^3} P^{\mathrm{lin}}(q) \left[ \frac{5}{6} + \frac{15}{8} S_2(\mathbf{q}, \mathbf{k} - \mathbf{q}) S_2(-\mathbf{q}, \mathbf{k}) - \frac{5}{4} S_2(\mathbf{q}, \mathbf{k} - \mathbf{q}) \right]
\end{align}
With the $F_2$, $G_2$ and $S_2$ kernels are defined as \cite{Peebles1975,Catelan1994}
\begin{align}
F_2(\mathbf{k}_1, \mathbf{k}_2) &= \frac{5}{7} + \frac{2}{7} \left( \frac{\mathbf{k}_1 \cdot \mathbf{k}_2}{k_1 k_2} \right)^2 + \frac{\mathbf{k}_1 \cdot \mathbf{k}_2}{2} \left( \frac{1}{k_1^2} + \frac{1}{k_2^2} \right), \\
G_2(\mathbf{k}_1, \mathbf{k}_2) &= \frac{3}{7} + \frac{\mathbf{k}_1 \cdot \mathbf{k}_2}{2} \left( \frac{1}{k_1^2} + \frac{1}{k_2^2} \right) + \frac{4}{7} \left( \frac{\mathbf{k}_1 \cdot \mathbf{k}_2}{k_1 k_2} \right)^2, \\
S_2(\mathbf{k}_1, \mathbf{k}_2) &= \left( \frac{\mathbf{k}_1 \cdot \mathbf{k}_2}{k_1 k_2} \right)^2 - \frac{1}{3}.
\end{align}

Finally, the 1-loop SPT predictions for the $P_{\delta\delta}^{\text{SPT}}$, $P_{\delta\theta}^{\text{SPT}}$ and $P_{\theta\theta}^{\text{SPT}}$ fields are given by
\begin{equation}
    P_{ij}^{\text{SPT}}(k) = P^{\text{lin}}(k) + 2P_{ij}^{(13)}(k) + P_{ij}^{(22)}(k). 
\end{equation}

The relative velocity effect in the power spectrum is sourced by the $G_u$ kernel defined as \cite{Yoo2011,Slepian2015},
\begin{align}
G_u(\mathbf{k}_1, \mathbf{k}_2) &= -T_v(k_1) T_v(k_2).
\end{align}

The integral form of the 1-loop real space addition due to relative velocity effects are
\begin{align}
\label{eq:real_relvel}
P_{\text{adv}|\delta}(k) &= \frac{4}{3} T_v(k) k P_{\text{m}}^{\text{lin}}(k) \int \frac{k \, dk}{2\pi^2} T_v(k) P_{\text{lin}}(k), \\
P_{\delta|v^2}(k) &= 4 \int \frac{d^3\mathbf{q}}{(2\pi)^3} P_{\text{m}}^{\text{lin}}(q) P_{\text{m}}^{\text{lin}}(k-q) F_2(\mathbf{q}, \mathbf{k}-\mathbf{q}) G_u(\mathbf{q}, \mathbf{k}-\mathbf{q}) \mu(\mathbf{q}, \mathbf{k}-\mathbf{q}),\\
P_{\delta^2|v^2}(k) &= 2 \int \frac{d^3\mathbf{q}}{(2\pi)^3} P_{\text{m}}^{\text{lin}}(q) \Big[ P_{\text{m}}^{\text{lin}}(k-q) \mu(\mathbf{q}, \mathbf{k}-\mathbf{q}) G_u(\mathbf{q}, \mathbf{k}-\mathbf{q}) \nonumber\\ 
&\qquad + P_{\text{m}}^{\text{lin}}(q) G_u(\mathbf{q}, \mathbf{q}) \Big], \\
P_{s^2|v^2}(k) &= 2 \int \frac{d^3\mathbf{q}}{(2\pi)^3} P_{\text{m}}^{\text{lin}}(q) \Big[ P_{\text{m}}^{\text{lin}}(k-q) S_2(\mathbf{q}, \mathbf{k}-\mathbf{q}) \mu(\mathbf{q}, \mathbf{k}-\mathbf{q}) G_u(\mathbf{q}, \mathbf{k}-\mathbf{q}) \nonumber\\ &\qquad + \frac{2}{3} P_{\text{m}}^{\text{lin}}(q) G_u(\mathbf{q}, \mathbf{q}) \Big],\\
P_{v^2|v^2}(k) &= 2 \int \frac{d^3\mathbf{q}}{(2\pi)^3} P_{\text{m}}^{\text{lin}}(q) \Big[ P_{\text{m}}^{\text{lin}}(k-q) \mu^2(\mathbf{q}, \mathbf{k}-\mathbf{q}) G_u^2(\mathbf{q}, \mathbf{k}-\mathbf{q}) \nonumber\\ & \qquad - P_{\text{m}}^{\text{lin}}(q) G_u^2(\mathbf{q}, \mathbf{q}) \Big],\\
P_{\delta \mid \delta_{bc}}(k) &= T_{bc}(k)\,P_{\mathrm{lin}}(k), \\
P_{\delta \mid \theta_{bc}}(k) &= \frac{\sigma_{vbc}}{H_0}\,T_v(k)\,k\,P_{\mathrm{lin}}(k).
\end{align}

The redshift space relative velocity contributions read
\begin{align}
P_{\delta|v^2 v_{||}}(k) &= \frac{2}{3} T_v(k) k P_{\text{lin}}(k)\int \frac{k \, dk}{2\pi^2} T_v(k) P_{\text{lin}}(k) = \frac{1}{2} P_{\text{adv}|\delta}(k), \\
P_{v^2|v_{||}}(k) &= 2 \int \frac{d^3\mathbf{q}}{(2\pi)^3} \frac{k\mu - q}{\sqrt{k^2 - 2kq\mu + q^2}} P_{\text{lin}}(q) P_{\text{lin}}(k-q) \notag \\ &\qquad\quad\times G_2(\mathbf{q}, \mathbf{k}-\mathbf{q}) G_u(\mathbf{q}, \mathbf{k}-\mathbf{q}), \\
P_{\text{adv}|v_{||}}(k) &= -\frac{2}{3} T_v(k) k P_{\text{lin}}(k)\int \frac{k \, dk}{2\pi^2} T_v(k) P_{\text{lin}}(k) = -\frac{1}{2} P_{\text{adv}|\delta}(k), \\
P_{v^2|\delta v_{||}}(k) &= 2 \int \frac{d^3\mathbf{q}}{(2\pi)^3} \frac{k\mu(k\mu - q)}{q\sqrt{k^2 - 2kq\mu + q^2}} P_{\text{lin}}(q) P_{\text{lin}}(k-q) G_u(\mathbf{q}, \mathbf{k}-\mathbf{q}), \\
P_{v_{||}|v^2 v_{||}}(k) &= -\frac{4}{3} T_v(k) k P_{\text{lin}}(k)\int \frac{k \, dk}{2\pi^2} T_v(k) P_{\text{lin}}(k) = -P_{\text{adv}|\delta}(k), \\
P_{v^2|v_{||}^2}(k) &= I_1(k) + \mu^2 I_2(k)
\end{align}
where
\begin{align}
I_1(k) &= k^2 \int \frac{d^3\mathbf{q}}{(2\pi)^3} \frac{k^2(1-\mu^2)(q - k\mu)}{[k^2 - 2kq\mu + q^2]^{3/2}} G_u(\mathbf{q}, \mathbf{k}-\mathbf{q}) P_{\text{lin}}(q) P_{\text{lin}}(k-q), \\
I_2(k) &= k^2 \int \frac{d^3\mathbf{q}}{(2\pi)^3} \frac{k^2(2k^2\mu^2 - k(3\mu^3 + \mu)q + (3\mu^2 - 1)q^2)}{q [k^2 - 2kq\mu + q^2]^{3/2}}\notag \\ &\qquad\qquad \times G_u(\mathbf{q}, \mathbf{k}-\mathbf{q}) P_{\text{lin}}(q) P_{\text{lin}}(k-q).
\label{eq:redshift_relvel}
\end{align}

\begin{figure} [ht]
    \centering
    \includegraphics[width=\textwidth]{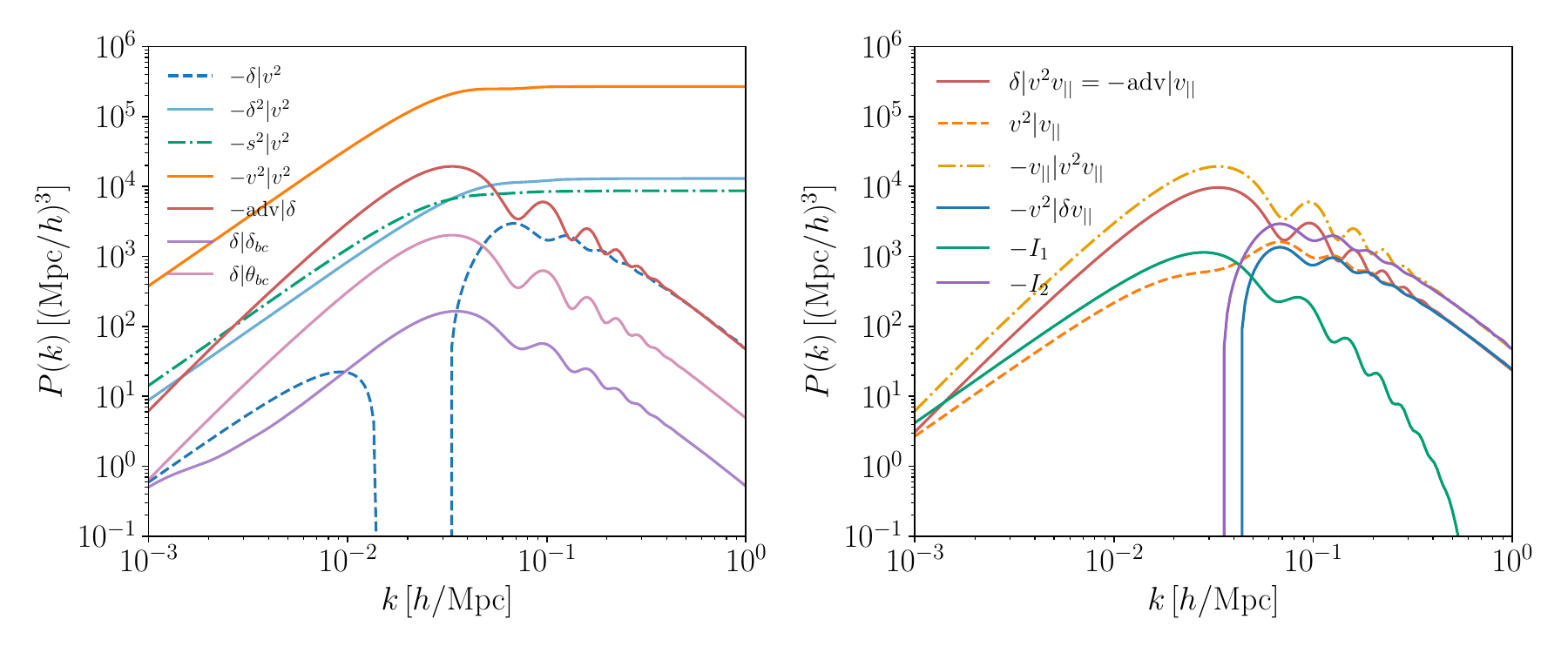}
    \caption{Comparison of the relative velocity contributions to the power spectrum model developed in \cite{Beutler2017}. We partially reproduce figure 1 of their paper to show the contributions to the power spectrum. The left panel shows the real space contributions, while the right panel has the RSD contributions.}
    \label{fig:integrals_pk}
\end{figure}

Figure \ref{fig:integrals_pk} shows the different relative velocity contributions to the power spectrum, which are defined in eq.~\ref{eq:real_relvel} - \ref{eq:redshift_relvel}. The left panel shows the real space contributions, while the right panel presents the RSD contributions. We observe the presence of oscillations in some of the relative velocity terms, which are out of phase with the BAO and overlap with the signal for the same range of wave-vectors. Those terms that present the oscillations are the most relevant ones when contributing to bias the $\alpha_{\rm iso}$ measurement.
\section{Modeling of the BAO in the Bispectrum}
\label{sec:appendix_B}
To extract the BAO signal from the bispectrum we use a model inspired by the tree-level expression of the redshift space bispectrum monopole \cite{Verde1998,Scoccimarro1999}
\begin{align}
\label{eq:bispectrumBAO}
    B^{(0)}(k_1,k_2,k_3;\alpha_{\rm iso})&=P(k_1)P(k_2)[ \mathcal{M}_1(\beta_F;k_1,k_2,k_3)F_2^{\rm eff}(k_1,k_2,k_3)\nonumber \\&+ \mathcal{M}_2(\beta_G;k_1,k_2,k_3)G_2^{\rm eff}(k_1,k_2,k_3)\nonumber \\&+\frac{1}{2}\mathcal{M}_1(\beta_F;k_1,k_2,k_3)(C_1S_2(k_1,k_2,k_3)+C_2)\nonumber \\&+\mathcal{M}_3(\beta_\mu;k_1,k_2,k_3) ] +{\rm cyc.}
\end{align}
The $\mathcal{M}_i$ functions account for the bias and the effect of redshift space distortions of the broadband. Their form reads 
\begin{align}
\mathcal{M}_1(\beta, k_1, k_2, k_3) &= \frac{2}{15} \left( 15 + 10\beta + \beta^2 + 2\beta^2 x^2 \right)\\
\mathcal{M}_2(\beta, k_1, k_2, k_3) &= \frac{2\beta}{105k_3^2} \big( 35k_1^2 + 28\beta k_1^2 + 3\beta^2 k_1^2 + 35k_2^2 + 28\beta k_2^2 + 3\beta^2 k_2^2 + 70k_1 k_2 x \nonumber\\
&\quad + 84\beta k_1 k_2 x + 18\beta^2 k_1 k_2 x + 14\beta k_1^2 x^2 + 12\beta^2 k_1^2 x^2 + 14\beta k_2^2 x^2 \nonumber\\
&\quad + 12\beta^2 k_2^2 x^2 + 12\beta^2 k_1 k_2 x^3 \big)
\end{align}
\begin{align}
\mathcal{M}_3(\beta, k_1, k_2, k_3) &= \frac{\beta}{315 k_1 k_2} \big( 210k_1 k_2 + 210\beta k_1 k_2 + 54\beta^2 k_1 k_2 + 6\beta^3 k_1 k_2 + 105k_1^2 x \nonumber\\
&\quad + 189\beta k_1^2 x + 99\beta^2 k_1^2 x + 15\beta^3 k_1^2 x + 105k_2^2 x + 189\beta k_2^2 x + 99\beta^2 k_2^2 x \nonumber\\
&\quad + 15\beta^3 k_2^2 x + 168\beta k_1 k_2 x^2 + 216\beta^2 k_1 k_2 x^2 + 48\beta^3 k_1 k_2 x^2 \nonumber\\
&\quad + 36\beta^2 k_1^2 x^3 + 20\beta^3 k_1^2 x^3 + 36\beta^2 k_2^2 x^3 + 20\beta^3 k_2^2 x^3 + 16\beta^3 k_1 k_2 x^4 \big).
\end{align}

For this analysis we substitute the $F_2$ and $G_2$ kernels by effective analogues with free fitting parameters calibrated using N-body simulations.  They have been shown to improve the performance of the tree-level form of the matter bispectrum up to scales of $k \simeq 0.2\,h{\rm Mpc}^{-1}$ for a wide range of redshifts, $z \leqslant 1.5$ \cite{hgm2012b}, \cite{hgm2014b}, 

\begin{align}
F_2^{\text{eff}}(\mathbf{k}_i, \mathbf{k}_j) &= \frac{5}{7}a(n_i, k_i; \mathbf{a}^\mathcal{F})a(n_j, k_j; \mathbf{a}^\mathcal{F}) + \frac{1}{2}\cos(\theta_{ij}) \left(\frac{k_i}{k_j} + \frac{k_j}{k_i}\right) b(n_i, k_i; \mathbf{a}^\mathcal{F})b(n_j, k_j; \mathbf{a}^\mathcal{F}) \nonumber \\
&+ \frac{2}{7}\cos^2(\theta_{ij})c(n_i, k_i; \mathbf{a}^\mathcal{F})c(n_j, k_j; \mathbf{a}^\mathcal{F}),  \\
G_2^{\text{eff}}(\mathbf{k}_i, \mathbf{k}_j) &= \frac{3}{7}a(n_i, k_i; \mathbf{a}^\mathcal{G})a(n_j, k_j; \mathbf{a}^\mathcal{G}) + \frac{1}{2}\cos(\theta_{ij}) \left(\frac{k_i}{k_j} + \frac{k_j}{k_i}\right) b(n_i, k_i; \mathbf{a}^\mathcal{G})b(n_j, k_j; \mathbf{a}^\mathcal{G}) \nonumber \\
&+ \frac{4}{7}\cos^2(\theta_{ij})c(n_i, k_i; \mathbf{a}^\mathcal{G})c(n_j, k_j; \mathbf{a}^\mathcal{G}), 
\end{align}

with the functions $a$, $b$ and $c$ defined as,
\begin{align}
a(n, k, \mathbf{a}) &= \frac{1 + \sigma_8^{a_6}(z) [0.7 Q_3(n)]^{1/2} (q a_1)^{n+a_2}}{1 + (q a_1)^{n+a_2}},  \\
b(n, k, \mathbf{a}) &= \frac{1 + 0.2 a_3 (n+3) (q a_7)^{n+3+a_8}}{1 + (q a_7)^{n+3.5+a_8}},\\
c(n, k, \mathbf{a}) &= \frac{1 + 4.5 a_4 / [1.5 + (n+3)^4] (q a_5)^{n+3+a_9}}{1 + (q a_5)^{n+3.5+a_9}}, 
\end{align}

where $q \equiv k/k_{\text{nl}}$ with $k_{\text{nl}}(z)$ a characteristic scale defined as,$\frac{k_{\text{nl}}(z)^3 P^{\text{lin}}_{\text{nw}}(k_{\text{nl}}, z)}{2\pi^2} \equiv 1$ and $n$ is the slope of the smoothed linear power spectrum,
$n(k) \equiv \frac{d \log P^{\text{lin}}_{\text{nw}}(k)}{d \log k}$.
Here $Q_3(n)$ is defined as, $Q_3(n) \equiv \frac{4 - 2^n}{1 + 2^{n+1}}$ and $\mathbf{a} = \{a_1, \dots, a_9\}$, is a set of nine free parameters to be fit by comparison to N-body simulations.
For ${F}_2^{\text{eff}}$ and ${G}_2^{\text{eff}}$, these parameters are \cite{hgm2012b,hgm2014b},
\begin{align*}
    a^\mathcal{F} = [0.484,0.392,0.128,3.740,1.013,-0.722,-0.849,-0.575,-0.926]\\
a^\mathcal{G} = [3.599,-3.588,5.022,-3.879,0.336,-3.104,0.518,7.431,-0.484]
\end{align*}

In this analysis, for the power spectrum $P(k)$ in equation \ref{eq:bispectrumBAO}, we apply the isotropic template to extract the BAO signal in the power spectrum monopole ($\ell=0$) \cite{eBossBAO},
\begin{equation}
    P(k;\alpha_{\rm iso})=\left[B_bP^{(\rm sm)}_{\rm{lin}}(k)+\sum^n_{i=1}A_{b,i}^{(0)}k^{2-i}\right]\times\left\{1+[\mathcal{O}_{\rm{lin}}(k/\alpha_{\rm iso})-1] e^{-\frac{1}{2}k^2(\Sigma^B_{\rm nl} )^2}\right\}.
\end{equation}
Note that here the $A_{b,i}^{(0)}$ are different the ones used for the power spectrum multipoles.
With this phenomenological approach we can capture the isotropic BAO component in the bispectrum monopole signal.

To validate the method, we fit this model to the Quijote mocks, described in section \ref{sec:covariance}. Our goal is to quantify the precision and accuracy of the recovery of $\alpha_{\rm iso}$ in the bispectrum monopole, as well as its consistency with the pre-reconstruction and post-reconstruction measurements from the power spectrum.
Ultimately, we want to provide a recipe to perform a joint MCMC measurement of the BAO parameters using both the power spectrum and bispectrum monopole.

We run an MCMC fit with $A_{b,i}^{(0)}$, $C_i$ and $\beta_i$ as free parameters to absorb the shape of the broadband, also marginalizing over $\Sigma^B_{\rm nl}$ and $B_b$ for the bispectrum. In this case the physical parameter is $\alpha_{\rm iso}$. 
We fit the model on the mean of 15000 independent realizations of the fiducial cosmology at redshift $z=0.5$. We perform the fit for the range $0.02\le k\,(h/{\rm Mpc})\le0.3$. \CH{We extend our bispectrum fits to relatively high values of $k$, despite the bispectrum entering the non-linear regime at lower $k$ than the power spectrum, and consider a range comparable to that typically adopted in power-spectrum analyses. Our goal is to exploit the widest possible range over which the BAO signal remains sufficiently undamped, while restricting the number of triangle configurations to a computationally manageable sample, particularly for the covariance estimation. Although our model is based on the tree-level expression for the bispectrum monopole and therefore does not fully capture its nonlinear evolution or broadband shape at these scales, we allow sufficient freedom to marginalize over the broadband contribution. This enables us to isolate and robustly extract the BAO signal without requiring an accurate description of the full bispectrum shape.}

\begin{figure} [ht]
    \centering
    \includegraphics[width=\textwidth]{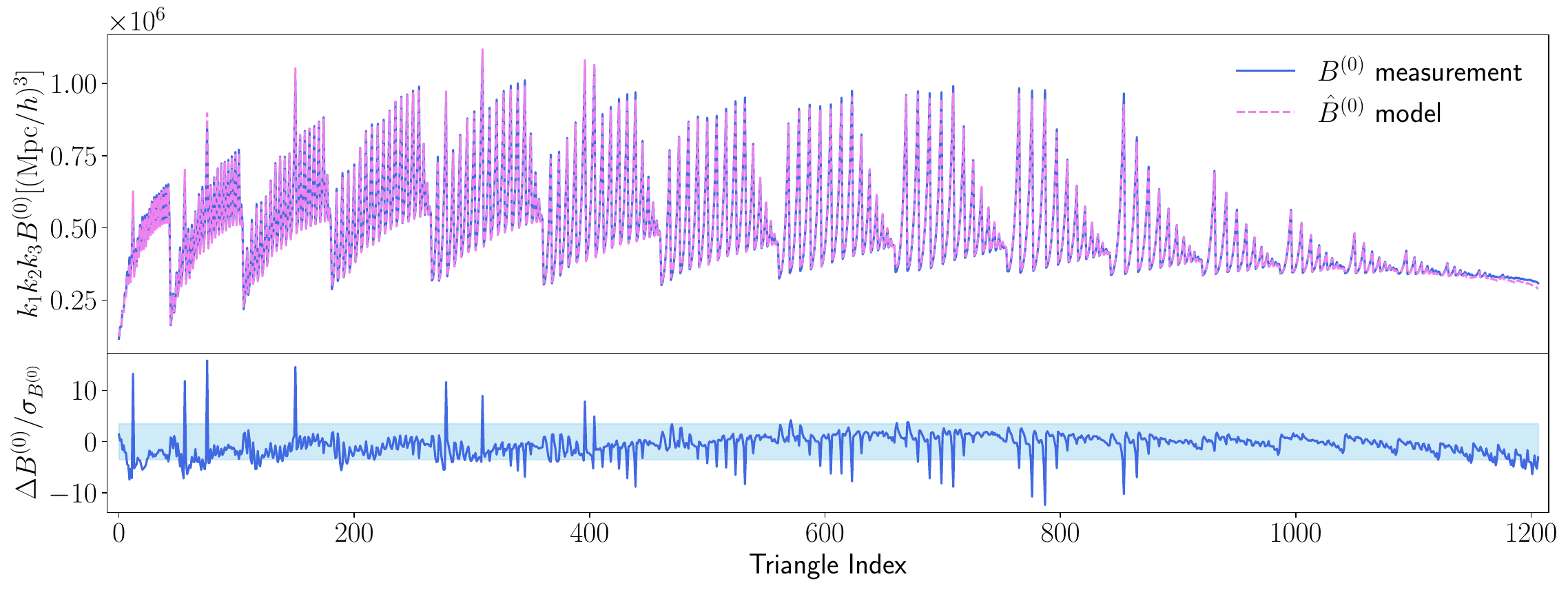}
    \caption{Performance of the BAO bispectrum model. The top panel shows the measured bispectrum, obtained from the mean of 15000 realizations of N-body cubic boxes. The overlapped dashed pink line shows the resulting best-fit model from the MCMC fit on the data. The bottom panel shows the residuals of the fit, computed with an error corresponding to 500($h^{-1}$Gpc)$^3$. \CH{A blue band marks the 3$\sigma$ region.}}
    \label{fig:B0_model}
\end{figure}

Figure \ref{fig:B0_model} shows the best-fitting model for the mean of the mocks. The top panel shows the measured monopole, in blue, obtained using \textsc{Rustico}\footnote{\href{https://github.com/hectorgil/Rustico}{https://github.com/hectorgil/Rustico}}. Overlaid in pink we show the best-fitting model, obtained using \textsc{cobaya}\footnote{\href{https://github.com/CobayaSampler/cobaya}{https://github.com/CobayaSampler/cobaya}} \cite{cobaya}. The $x$ axis displays the triangle index ordered by ascending value of $k_1$. The bottom panel shows the residuals of the fit, corresponding to a volume of 500($h^{-1}$Gpc)$^3$. \CH{We overlay a blue band, corresponding to the 3$\sigma$ region.} We can see the model properly captures the overall shape of the bispectrum monopole. The residuals stay within 3$\sigma$, except for a handful of triangle configurations which have predominantly been identified as flattened ($k_1=k_2, \,k_3=k_1+k_2,\, \cos\theta_{2,3}=-0.99$). A fit performed omitting these configurations verifies they do not introduce any biases in the measurement. 
\CH{The theoretical model begins to show deviations at small physical scales (high values of $k$). In this regime, our tree-level based model becomes less precise across all triangle geometries, as  non-linearities begin to introduce shape-dependent effects.
However, because the extraction relies on isolating the relative wiggles rather than fitting the exact full shape of the broadband continuum, the BAO measurement remains unaffected by this high-k trend.
A secondary systematic can arise from our evaluation method. The most robust way to compare the theoretical model with the measurements is to average the model predictions over all triangle configurations within each bin. Since this procedure is computationally expensive, we instead evaluate the bispectrum model for a single effective triangle in each bin, defined by the average values of $(k_1,k_2,k_3)$ over the configurations contained in that bin \cite{Sefusatti2010}. This approximation can introduce non-negligible deviations, particularly at larger scales where bins contain a smaller number of triangle configurations.
Ultimately, because our analysis focuses exclusively on extracting the BAO signal rather than modeling the full measurement shape, we expect both the small-scale non-linear deviations and the effective-triangle approximations to be subdominant. Consequently, these systematic limitations are not expected to introduce a significant bias in the recovered $\alpha_{iso}$.}

\section{Validation of the BAO Extraction from the Bispectrum Monopole}
\label{sec:appendix_C}
In this section we validate the performance of the model presented in app.~\ref{sec:appendix_B} for measuring the BAO. We extract the isotropic BAO signal from the monopole of the mean of 15000 realizations of N-body simulations, described in sec.~\ref{sec:covariance}. We also perform the fit to 500 independent realizations to estimate the variance of the measure.

\begin{figure} [ht]
    \centering
    \includegraphics[width=0.51\textwidth]{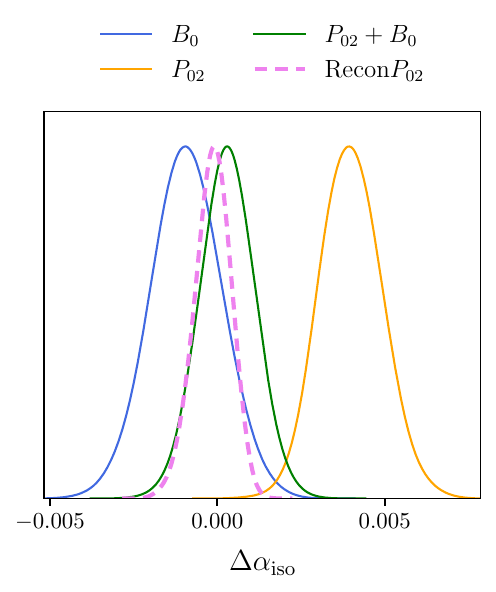}
    \caption{Comparison of the 1D posteriors of $\alpha_{\rm iso}$ obtained from performing MCMC fits on the power spectrum multipoles ($P_{02}$, orange), the bispectrum monopole ($B_0$, blue), a joint fit using both ($P_{02} + B_0$, green). For comparison, the constraint obtained from the reconstructed power spectrum is also shown (Recon$P_{02}$, pink). \CH{All errors assume a survey volume of 500($h^{-1}$Gpc)$^3$. The $\sim5\%$ shift on the pre-reconstruction $\alpha_{\rm iso}$ form fitting the power spectrum only is as expected and consistent with other finding in the literature (see text for details)}.}
    \label{fig:alpha_posteriors_comparison}
\end{figure}

Figure \ref{fig:alpha_posteriors_comparison} displays the shift in $\alpha_{\rm iso}$, recovered using different data-vectors. \CH{The covariances assumed correspond to the statistical error of a survey volume of 500($h^{-1}$Gpc)$^3$.} We show in orange the result from the pre-reconstructed power spectrum monopole and quadrupole ($P_{02}$) measurement, obtaining a shift of $\Delta\alpha_{\rm iso}\sim 0.5\%$ expected from non-linearities introduced by galaxy bulk flows and the gravitational collapse of large cluster \cite{Eisenstein_2007_BAO,Blas2016,Sherwin2012}. In blue we show the bispectrum only measurement ($B_0$). We obtain an unbiased measurement, with a comparable constraining power to that of the $P_{02}$ measurement. In green, the joint analysis, using both the power spectrum multipoles and the bispectrum monopole (Joint $P_{02} + B_0$). The joint fit is able to recover an unbiased measurement of $\alpha_{\rm iso}$, with $\Delta\alpha_{\rm iso}\approx 0$. It also increases the constraining power by $22\%$ from the power spectrum only one, showing the bispectrum monopoles carries information that complements that of the power spectrum first two multipoles. Finally, we plot for comparison the obtained posterior for the post-reconstructed field, measured using the power spectrum only (Recon$P_{02}$). We see the constraining power of the post-reconstructed measurement is increased $\sim50\%$ from the pre-reconstructed $P_{02}$, and about 30$\%$ more than the joint analysis $P_{02} + B_0$. When fitting the post-reconstructed power spectrum monopoles together with the pre-reconstructed bispectrum monopole we have observed no significant increase in the constraining power, suggesting the information contained in the bispectrum monopole is recovered in the post reconstructed power spectrum.

\begin{figure} [ht]
    \centering
    \includegraphics[width=\textwidth]{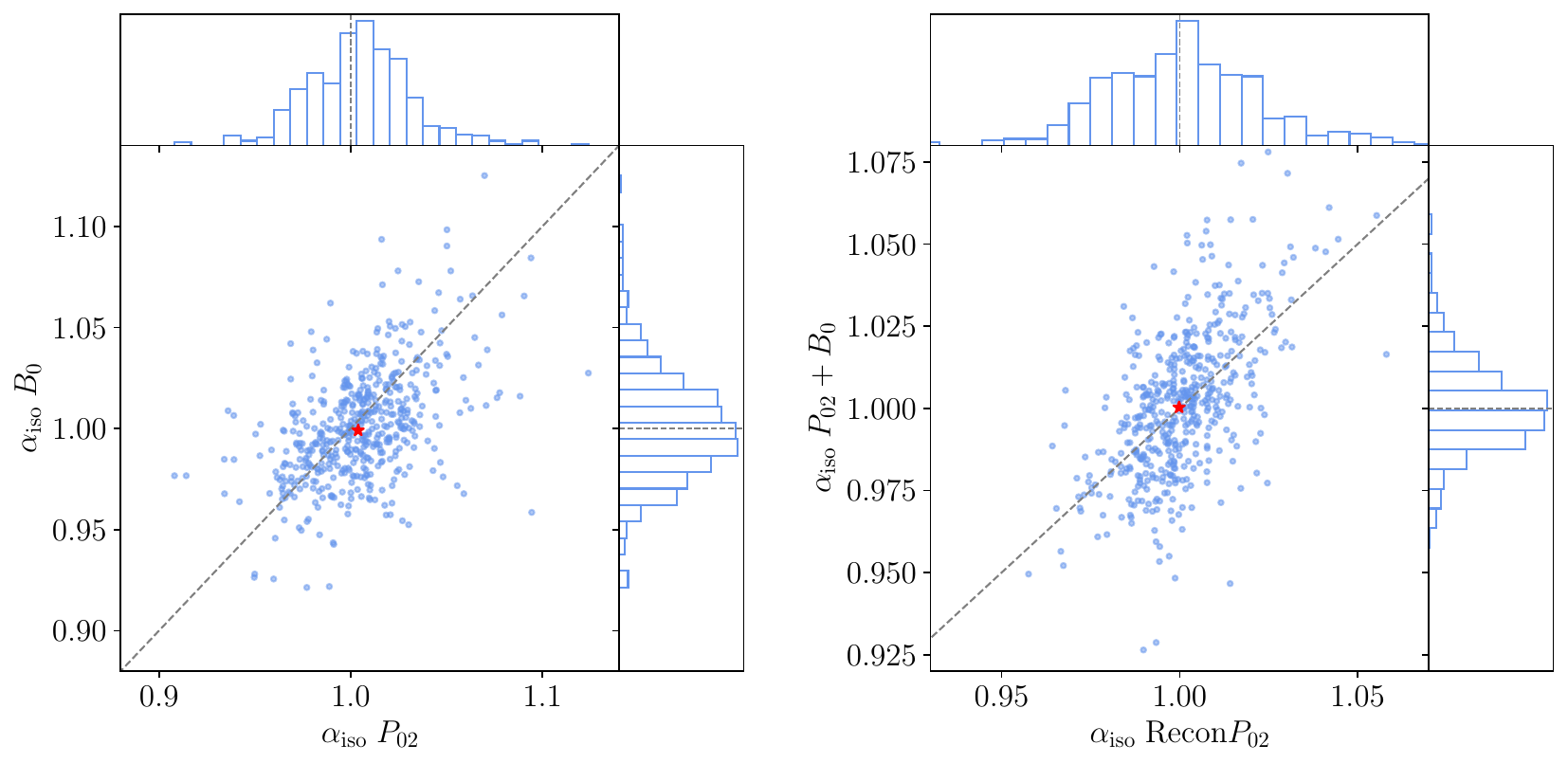} 
    \caption{Scatter of the best fit measurements of $\alpha_{\rm iso}$. The performance of 500 individual realizations of the Quijote mocks are shown in blue, while the red star corresponds to the fit obtained from the mean of 15000 realizations. Left panel shows the correlation between the power spectrum and the bispectrum only measurements of $\alpha_{\rm iso}$. Right panel shows the correlation between the joint $P_{02} + B_0$ fit and the post-reconstruction power spectrum fit for the recovered $\alpha_{\rm iso}$.}
    \label{fig:alpha_mock_comparison}
\end{figure}

\begin{table}[ht]
\centering
\begin{tabular}{|c|c|c|c|c|}
\hline
Catalog / Statistic & $<\Delta\alpha_{\rm iso}>$ [$\%$] & $<\sigma_{\rm iso}>[\%]$ & $<\Delta\alpha_{\rm AP}>$ [$\%$] & $<\sigma_{\rm AP}>[\%]$ \\ \hline
\hline
Pre-recon $P_{02}$         & 0.51 & 2.80 & 0.26 & 10.89 \\ \hline
Pre-recon $B_0$            & 0.11 & 2.68 &   -   &   -   \\ \hline
Post-recon $P_{02}$        & 0.08 & 1.42 & 0.14 & 4.69 \\ \hline
Pre-recon $P_{02} + B_0$   & 0.15 & 2.17 & 0.30 & 8.97 \\ \hline
Post-recon $P_{02} + B_0$  & 0.06 & 1.34 & 0.25 & 4.58 \\ \hline
\end{tabular}
\caption{Table summarizing the mean and dispersion of the recovered $\alpha_{\rm iso}$ and $\alpha_{\rm AP}$, as well as their statistical error, $\sigma_{\rm iso}$ and $\sigma_{\rm AP}$ respectively. Each row corresponds to a mentioned catalog combination, pre-recon $P_{02}$, $B_0$ and joint $P_{02}+B_0$; as well as post-recon $P_{02}$ and joint $P_{02} + B_0$.}
\label{tab:alphas}
\end{table}

Figure \ref{fig:alpha_mock_comparison} shows in blue the dispersion of the measurement of $\alpha_{\rm iso}$ for 500 independent realizations of the Quijote mocks. The measurement on the mean of 15000 realizations is shown as a red star. The left panel shows the power spectrum and bispectrum measurements of $\alpha_{\rm iso}$ are consistent. The right panel shows that the dispersion of our joint $P_{02} + B_0$ measurements is higher than the post-reconstructed measurements, corresponding to tighter constraints for the latter. 

A summary of the fits performed can be found in Table~\ref{tab:alphas}. We report the mean value of the shift in $\alpha_{\rm iso}$ and $\alpha_{\rm AP}$, as well as their dispersion, $\sigma_{\rm iso}$ and $\sigma_{\rm AP}$ respectively. We compare the results for all the mentioned catalog combinations, pre-recon $P_{02}$, $B_0$ and joint $P_{02} + B_0$; as well as post-recon $P_{02}$ and joint $P_{02} + B_0$. \CH{The reported errors correspond to the statistical error of a single realization of 1($h^{-1}$Gpc)$^3$.}

With this study, we validate the methodology to extract the isotropic BAO signal from the bispectrum monopole. We show it is an independent unbiased measurement that can also complement that of the power spectrum to improve the constraining power of the technique for the pre-reconstructed field, as well as in combination with the post-reconstruction measurement.

\bibliographystyle{JHEP}
\bibliography{references}
\end{document}